\documentclass[aps,prb,twocolumn,superscriptaddress,showpacs]{revtex4-1}
\usepackage[utf8]{inputenc}
\usepackage{amsmath}
\usepackage{amsfonts}
\usepackage{amssymb}
\usepackage{graphics}
\usepackage{tabularx}
\usepackage{mathtools}
\usepackage{float}
\begin{document}
\title{Delocalization of charge and current in a chiral quasiparticle wave packet}
\author{Subhajit Sarkar}
\email{sbhjt72@gmail.com, subhajit@iopb.res.in}
\affiliation{Institute of Physics, P.O.: Sainik School, Bhubaneswar 751005, Odisha, India.}
\date{\today}
\begin{abstract}\label{abs}
A chiral quasi-particle wave packet (c-QPWP) is defined as a conventional superposition of chiral quasi-particle states corresponding to an interacting electron system in two dimensions (2D) in the presence of Rashba spin-orbit coupling (RSOC). I investigate its internal structure via studying the charge and the current densities within the first order perturbation in the electron-electron interaction. It is found that the c-QPWP contains a localized charge which is less than the magnitude of the bare charge and the remaining charge resides at the system boundary. The amount of charge delocalized turns out to be inversely proportional to the degenerate Fermi velocity $v_0 ( = \sqrt{\alpha^2 + 2\mu / m})$ when RSOC (with strength $\alpha$) is weak, and therefore externally tunable. For strong RSOC, the magnitudes of both the delocalized charge and the current further strongly depend on the direction of propagation of the wave packet. Both the charge and the current densities consist of an anisotropic $r^{-2}$ tail away from the center of the wave packet. Possible implications of such delocalizations in real systems corresponding to 2D semiconductor heterostructure are also discussed within the context of particle injection experiments.
\end{abstract}
\maketitle
\section{Introduction}
The interplay of the chirality and the electron-electron (e-e) interaction is a very important issue from both fundamental and applied perspectives in many-body quantum systems. Bare electron develops chirality when its spin ($\sigma$) and momentum ($\mathbf{p}$) get locked because of the presence of spin-orbit (SO) coupling \cite{SDS, HZ}. Moreover, in an interacting electron system the presence of SO coupling brings an additional energy scale, apart from the Fermi energy and the Coulomb energy which were already present. The analysis of different phases in the interacting electron systems usually starts from a Fermi liquid theoretic point of view \cite{AGD, POM}. In the presence of SO coupling the interplay of all the above mentioned energy scales lead to the formation of new phases of matter \cite{CSG, JT, WZ, WSF, CM, BRK}. In this regard a theory of the chiral Fermi liquid (CFL) has been put forward recently along the line of the conventional Landau Fermi liquid theory by focusing on the presence of Rashba SO coupling \cite{ARM}. The central pillar in this CFL theory is the existence of chiral quasi-particles  which are valid only near the Fermi surfaces of the respective Rashba sub-bands \cite{ARM}.
\paragraph*{}
The Landau Fermi liquid theory is formulated in terms of the distribution function of quasi-particles $n(\mathbf{k,\, r})$, and this can be obtained from the well established microscopic calculations \cite{NL, AGD}. This distribution function is generally considered as a semi-classical quasi-particle wave packet (QPWP) of mean momentum $\mathbf{k}$, mean position $\mathbf{r}$, and charge $e$ (being equal to the charge of the bare particle). However, it has been shown that the QPWP develops a non-trivial internal structure because of the electron-electron interaction. This internal structure leads to the delocalization of charge and current in the QPWP state \cite{HK}.  In a spin-$\frac{1}{2}$ Landau quasi-particle wave packet (Landau-QPWP) with spin $\sigma = \uparrow$, the charge density consists of a localized (spherically symmetric) part corresponding to a charge $e^{\prime}$ such that $\frac{e^{\prime}}{e} \neq 1$, and the rest of the charge $(1 - \frac{e^{\prime}}{e}) < 1$, gets delocalized and uniformly distributed at the surface of the large volume \cite{HK}. Moreover, the  Landau-QPWP contains a localized spin $\sigma^{\prime} > \frac{1}{2}$, leading to the spin-charge separation in the Landau-QPWP \cite{OH}. The bare particle wave packet (in the absence of electron-electron interaction) on the other hand is structure less  with a charge equals unity ($\frac{e^{\prime}}{e}=1$) and spin of magnitude 1/2 localized within the spatial spread of the wave packet. It has been pointed out that because of this non-trivial internal structure of a QPWP in the presence of e-e interaction, the above mentioned distribution function $n(\mathbf{k,\, r})$ can't be interpreted as a QPWP \cite{HK, OH}.
\paragraph*{}
 The concept of QPWP is important in the tunnelling experiments in relation to reflection and transmission through a barrier \cite {BOC, MAR}. Experimentally it has been found that there exists a finite probability of finding both the electrons on the same side of the barrier when those two are injected from two different sources separated by the barrier \cite{BOC}. This phenomenon has been attributed as due to the fundamental wavepacket nature of the electron quasi-particles \cite{MAR}.
 \paragraph*{}
 Furthermore, in real systems the SO coupling remains an important character \cite{Manc, Con, SO}. In non-centrosymmetric semiconductors bulk SO coupling becomes odd in electron's momentum and this is known as the Dresselhaus coupling \cite{Dres}. In two-dimensional (2D) semiconductor heterostructures with structural inversion asymmetry the SO coupling becomes linear in electron's momentum and the corresponding SO coupling is well known as the Rashba SO coupling (RSOC) \cite{Ras1, Vasko, Bych, Wink}. Systems with RSOC have been investigated quite extensively, even at the single particle level, to uncover appearances of rich variety of exotic quantum phases \cite{Manc}. In particular, in 2D heterostructures the experiments are usually performed in a well controllable manner and the strength of the SOC can be tuned externally \cite{Hwang, Manc}.  Moreover, the interplay of SO coupling and electron-electron interaction also bring unconventional long range order in the system and interestingly enough, the SO coupling itself gets renormalized by the momentum dependent screened Coulomb interaction \cite{Kim, Set, Tada, Yan1, Taki, Yan2, Tada1, Taki1, Yoko, Tada2, Tada3}. However, the quasi-particle properties largely remain unaffected by the interplay between them in the Fermi liquid state \cite{Chen, Chesi, Aasen, Sara}.
\paragraph*{}
 In view of the importance of both the presence of SO coupling, and the wave packet nature of the quasi-particles in the real systems, in this article I consider the issue of delocalization of charge and current in the chiral QPWP (c-QPWP) corresponding to the Fermi liquid in the presence of Rashba SO coupling which is relevant in the 2D electron liquid appearing at the inversion layer of semiconductor heterostructures \cite{ARM, Vignale}. To the best of my knowledge the effect of the SOC on the internal structure of the QPWP has not been investigated in the literature to date.
\paragraph*{}
In this article, I define a c-QPWP of specific chirality `$s$' corresponding to the CFL as a conventional superposition of chiral quasi-particle states. Here, I have investigated the important role played by the RSOC in the internal structure of such c-QPWP by studying the expectation values of the Fourier transform of the charge density operator, $\hat{n}(\mathbf{q})$ and current density operator, $\hat{j}(\mathbf{q})$ in the c-QPWP state in the limit $|\mathbf{q}|\rightarrow 0$. The important results obtained in this paper are as follows. It is found that both $\hat{n}(\mathbf{q})$ and $\hat{j}(\mathbf{q})$ are discontinuous at $\mathbf{q} = 0$ which signals to the fact that the charge and current associated with the c-QPWP are delocalized to infinity, i.e., to the boundary of a thermodynamically large system. This is because of the effect of the e-e interaction, as found in the case of Landau-QPWP \cite{HK, OH}. However, in the presence of SOC the magnitudes of the delocalized charge and current depend on the strength of the SO coupling and also on the direction of propagation of the wave packet. Both weak and strong SOC have been considered. Furthermore, in the present case, the fact that strength of the RSOC is externally tunable makes the magnitude of the delocalized charge and current externally tunable. The dependence of the delocalized charge and current on the strength of RSOC is expected to aid the experimental detection of the delocalization effect.  On the contrary, the case of a conventional Landau Fermi liquid lacks any tuning parameter similar to the strength of the RSOC. Therefore, the observations of the localized charge and current carried by the Landau-QPWP have been extremely difficult \cite{OH}.
\paragraph*{}
 The article is organized as follows. The the charge and current densities of a c-QPWP have been presented in section \ref{2}. In section \ref{3}, I calculate the amount of charge and current delocalized to the boundary and in \ref{3a}, I discuss some experimental implications. The results are discussed in section \ref{4} and calculational details are presented in the Appendices.
\section{Charge and current of a chiral quasi-particle wave packet: general formulation}\label{2}
In this article, I consider the Hamiltonian corresponding to the 2D CFL which is described by,
\begin{eqnarray}\label{h}
H &=& H_0 + H_{int} \nonumber \\ &\text{with}& \nonumber \\ H_0 &=& \sum_{\mathbf{k}, \sigma, \sigma'} \Large[ c_{\mathbf{k}, \sigma}^{\dagger} \delta_{\sigma \sigma'}\frac{k^2}{2m} c_{\mathbf{k}, \sigma'} \nonumber \\ &+& c_{\mathbf{k}, \sigma}^{\dagger} \alpha(\tau^{x}_{\sigma \sigma'} k_{y} - \tau^{y}_{\sigma \sigma'} k_{x}) c_{\mathbf{k}, \sigma'}\Large],
\end{eqnarray}
where $\alpha$ is the strength of Rashba SO coupling (RSOC), $H_{int}$ is the conventional e-e interaction whose precise expression is given later in (\ref{int}), $k^2 = k_x^2 + k_y^2$, with $\hbar= e \text{(charge of electron)} =1$, and $m$ being the effective band mass \cite{ARM}. The quantities $\tau^{x}$ and $\tau^{y}$ are the $x$ and $y$ components of the Pauli matrices respectively and $\alpha$ is taken to be positive. The non-interacting part, $H_0$ is non-diagonal in the spin basis and is diagonalized using the following unitary matrix,
\begin{equation}\label{matrix}
U =\frac{1}{\sqrt{2}}\begin{pmatrix}
    1 & 1  \\
   - ie^{i\theta_{\mathbf{k}}} & ie^{i\theta_{\mathbf{k}}}
  \end{pmatrix},
\end{equation}
where $\theta_{\mathbf{k}} = \tan^{-1}\frac{k_y}{k_x}$ is the azimuth of $\mathbf{k}$. The diagonalized non-interacting Hamiltonian reads as,
\begin{equation}\label{h0}
H_0 = \sum_{\mathbf{k}, s} c_{\mathbf{k}, s}^{\dagger} \xi_{\mathbf{k},s} c_{\mathbf{k}, s}
\end{equation}
where $\xi_{\mathbf{k},s} = \frac{k^2}{2m} + s\alpha k$ and $s = \pm 1$ denotes the chirality or the winding direction of the spins around the Fermi surface  \cite{ARM}. This dispersion relation is shown in FIG. 1(a).
\begin{figure*}
\centering
\includegraphics[scale=0.20]{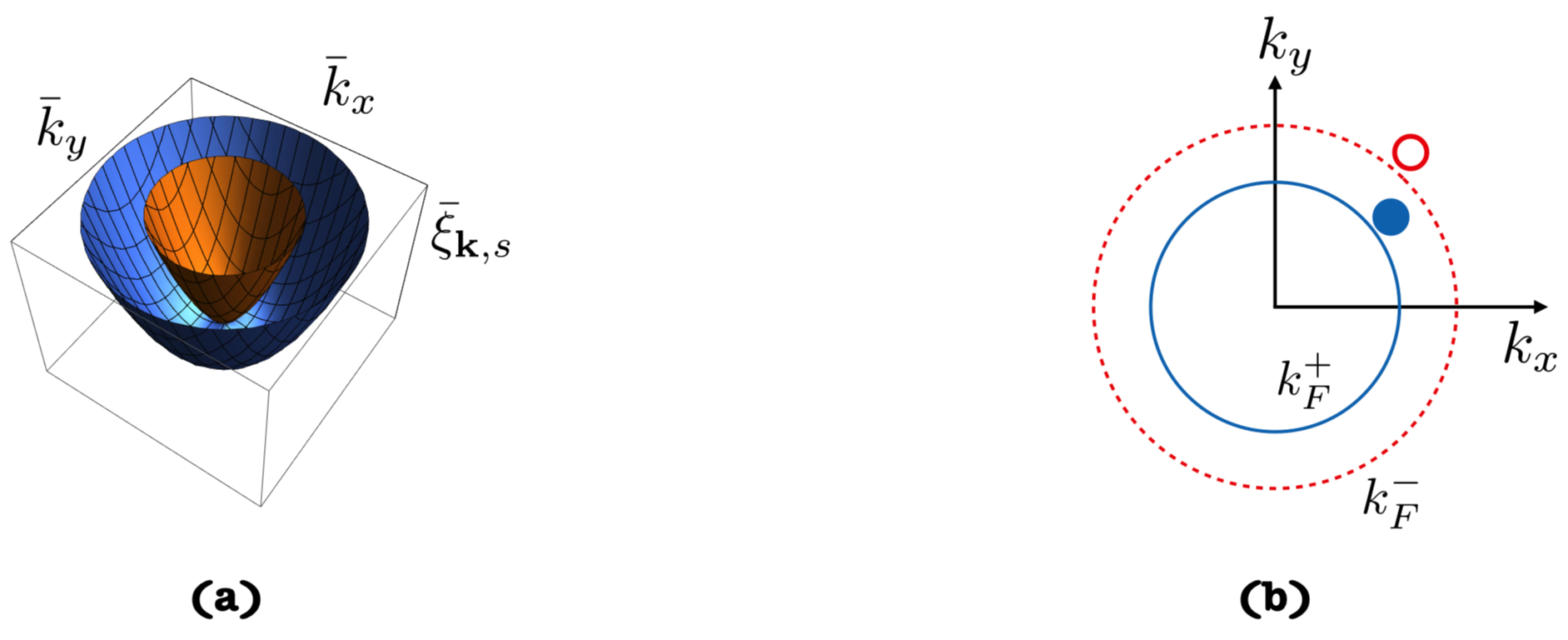}
\caption{(a) Dispersion relation $\bar{\xi}_{\mathbf{k},s}$ vs. $\bar{k}_x$ and $\bar{k}_y$, where z-axis represents $\bar{\xi}_{\mathbf{k},s} = \frac{\xi_{\mathbf{k},s}}{m\alpha^2}$, and $\bar{k}_x = \frac{k_x}{m\alpha}$ and $\bar{k}_y = \frac{k_y}{m\alpha}$. (b) Fermi surfaces are concentric circles, $k_{F}^{+}$ is the radius corresponding to $s = +1$, and $k_{F}^{-}$ is the radius corresponding to $s = -1$. Quasi-particle picture is valid only near the Fermi surface, quasi-particle being schematically shown as red circle and blue dot on the respective Fermi surfaces.}
\end{figure*} 
 The normalized plane wave states corresponding to the these chiral electrons are given by,
\begin{equation}\label{k}
|\mathbf{k}, s\rangle = \frac{1}{\sqrt{2}} \begin{pmatrix}
1 \\ -ise^{i \theta_{\mathbf{k}}}
\end{pmatrix} e^{i \mathbf{k} \cdot \mathbf{r}}, 
\end{equation}
where the 2D volume of the system $\Omega$ has been taken to be unity and standard periodic boundary conditions are assumed \cite{ARM}. For $\mu > 0$, the $\xi_{\mathbf{k},s} = \mu$ plane cuts the dispersion curves in such a way that the Fermi surfaces corresponding to both the Rashba sub-bands turns out to be concentric circles with radii $k_{F}^{s}$, as shown in FIG. 1(b). The Fermi momenta and the Fermi velocities of individual sub bands are given by, $k_{F}^{\pm} = m(v_0 \mp \alpha)$ and $v_0 = \sqrt{\alpha^2 +2\mu /m}$ respectively where $\mu$ is the chemical potential \cite{ARM}.  The strength of the RSOC is considered to be small thereby ensuring the Fermi velocities of two sub-bands to be same. However, in the case of strong RSOC the Fermi velocities corresponding to the individual sub-bands no longer remain degenerate and are self-consistently determined by the renormalized Fermi momenta \cite{ARM}. The ground state of the non-interacting system is the filled Fermi circle since the chiral electron also satisfy the Pauli exclusion principle. Therefore, the ground state is constructed by filling up all the single chiral particle states until the respective Fermi momentum, viz.,
\begin{equation}\label{gnd}
|F\rangle = \prod_{\mathbf{k'}<k_{F}^{+}}  \prod_{\mathbf{k''}<k_{F}^{-}}  c_{\mathbf{k'}, +}^{\dagger}c_{\mathbf{k''}, -}^{\dagger} |0\rangle ,
\end{equation}
$|0\rangle$ being the vacuum  \cite{ARM}. If there are $N$ chiral electrons, $N/2$ electrons with chirality `$+$' shall be within the Fermi 2-sphere (i.e., circle) of radius $k_{F}^{+}$, and rest $N/2$ electrons with chirality `$-$' shall be within the Fermi 2-sphere of radius $k_{F}^{-}$. Then one can find a one particle state by adding a bare electron of specified chirality above the corresponding Fermi sea, viz., $|\mathbf{k}, s \rangle = c_{\mathbf{k}, s}^{\dagger} |F\rangle$, for all $\mathbf{k} > k_{F}^{s}$. 
\paragraph*{}
$H_{int}$ further can expressed in the chiral basis by expanding the field operator $\hat{\psi}$ corresponding to the two body interaction as $\hat{\psi}(\mathbf{r}) = \sum_{\mathbf{k},s} c_{\mathbf{k},s} |\mathbf{k}, s\rangle$, and the form is given by \cite{Sara},
\begin{eqnarray}\label{int}
H_{int} &=& \frac{1}{2}\sum_{\mathbf{k_1, k_2 , p}, s_1 , s_2, s_3 , s_4} V_{s_1 , s_2, s_3 , s_4} (\mathbf{k_1 , k_2, p}) \nonumber \\ &c_{\mathbf{k_1 -p}, s_1}^{\dagger}& c_{\mathbf{k_2 +p}, s_2}^{\dagger} c_{\mathbf{k_2}, s_3} c_{\mathbf{k_1}, s_4},
\end{eqnarray}
where
\begin{eqnarray}\label{gamma}
& &V_{s_1 , s_2, s_3 , s_4} (\mathbf{k_1 , k_2, p}) = V(p) \frac{1}{4}[ 1 + s_1 s_4 e^{i( \theta_{\mathbf{k_1}} - \theta_{\mathbf{k_{1} - p}} )} \nonumber \\  &+& s_2 s_3 e^{i(\theta_{\mathbf{k_2}} - \theta_{\mathbf{k_{2} + p}})} + s_1 s_2 s_3 s_4 e^{i(\theta_{\mathbf{k_1}} - \theta_{\mathbf{k_{1} - p}}+\theta_{\mathbf{k_2}} - \theta_{\mathbf{k_{2} + p}})} ]. \nonumber \\
\end{eqnarray}
In this article, I consider $V(p) = \frac{2\pi}{\sqrt{|\mathbf{p}|^{2} + \delta^{2}}}$ as the Fourier transform of the Coulomb  (2D-projected) interaction in two dimensions (2D), where for the purpose it is sufficient to consider $\delta$ to be a term regularizing the Coulomb interaction so that $V(p)$ does not diverge as $p = |\mathbf{p}| \rightarrow 0$ \cite{Vignale, RC, RC1}. By switching on the interaction (represented by the above Hamiltonian (\ref{int})) adiabatically the one particle state $|\mathbf{k}, s \rangle$ can be evolved into a chiral quasi-particle state $|\psi_{\mathbf{k},s} \rangle$ \cite{ARM}. 
\paragraph*{}
In this paper, I have calculated the charge and current density of a c-QPWP, defined in section \ref{defn}, within the 1st order perturbation in the e-e interaction. For this purpose in the following I have re-written the charge and the current density operators in the chiral basis. 
\subsection{Charge and current density operators in the chiral basis}\label{operat}
The charge density operator $\hat{n}(\mathbf{q})$ is expressed in the chiral representation as,
\begin{equation}\label{charge}
\hat{n}(\mathbf{q}) = \sum_{\mathbf{k'},s, s'} c_{\mathbf{k'-q},s}^{\dagger} c_{\mathbf{k'},s'} \frac{1}{2}[1+ s s' e^{-i(\theta_{\mathbf{k'-q}} - \theta_{\mathbf{k'}})} ],
\end{equation}
where the charge $e = 1$, as I have already mentioned and $\mathbf{k'}$ in the sum is unrestricted. The charge-current or simply current density operator $\hat{j}(\mathbf{q})$ is obtained in two steps, viz., first current density operator is obtained in the Pauli basis, which is given by \cite{BND},
\begin{eqnarray}\label{curr}
\hat{\mathbf{j}}(\mathbf{q}) &=& \hat{\mathbf{j}}_{kin}(\mathbf{q}) + \hat{\mathbf{j}}_{SO}(\mathbf{q}) \nonumber \\ &=& \sum_{\mathbf{k'},\sigma, \sigma '} \frac{1}{m}(\mathbf{k'}-\frac{\mathbf{q}}{2}) \delta_{\sigma, \sigma '}c_{\mathbf{k'-q},\sigma}^{\dagger} c_{\mathbf{k'},\sigma '}  \nonumber \\ &+& \sum_{\mathbf{k'},\sigma, \sigma '} \alpha (\tau_{\sigma, \sigma '}^{x} \hat{y} - \tau_{\sigma, \sigma '}^{y} \hat{x}) c_{\mathbf{k'-q},\sigma}^{\dagger} c_{\mathbf{k'},\sigma '},
\end{eqnarray}
where $\mathbf{k'}$ in the sum is unrestricted. Because of the presence of spin-orbit coupling the above equation for the current density operator consists of two parts, viz., a kinetic part, $\hat{\mathbf{j}}_{kin}(\mathbf{q})$ and a spin-orbit (SO) part, $\hat{\mathbf{j}}_{SO}(\mathbf{q})$. It can be easily seen that the spir-orbit part of the current density operator is composed of components of in-plane spin-density operators, thereby signifying a spin-charge coupled transport \cite{BND}. Then in the second step, I rewrite the operator in the chiral basis, and the kinetic part and the spin-orbit part corresponding to the current density operator take the form,
\begin{eqnarray}\label{curr_chi}
& &\hat{\mathbf{j}}_{kin}(\mathbf{q})\nonumber \\ &=& \sum_{\mathbf{k'},s, s'} \frac{1}{m}(\mathbf{k'}-\frac{\mathbf{q}}{2}) \frac{1}{2} [1+ s s' e^{-i(\theta_{\mathbf{k'-q}} - \theta_{\mathbf{k'}})}]  c_{\mathbf{k'-q},s}^{\dagger} c_{\mathbf{k'},s'},   \nonumber  \\ \nonumber \\ & \text{and}& \nonumber \\ & &
\hat{\mathbf{j}}_{SO}(\mathbf{q})\nonumber \\ &=& \sum_{\mathbf{k'},s, s'} \frac{\alpha}{2} [s' e^{i\theta_{\mathbf{k'}}} (\hat{x} - i \hat{y}) + s e^{-i\theta_{\mathbf{k'-q}}} (\hat{x} + i \hat{y})] c_{\mathbf{k'-q},s}^{\dagger} c_{\mathbf{k'},s'} \nonumber \\
\end{eqnarray}
In the following the formal definition of a c-QPWP is introduced.
\subsection{Definition of the c-QPWP}\label{defn}
To investigate the internal structure of a c-QPWP, in this article I define a c-QPWP with an average momentum $\mathbf{k_0}$ (propagating in the direction $\hat{\mathbf{k}}_0$) and chirality `$s$', as a superposition of the chiral quasi-particle states,
\begin{equation}\label{qpwp}
|\Psi_{\mathbf{k_0}, s} \rangle = \sum_{\mathbf{k}, |\mathbf{k}|\geq k_{F}^{s}} A_{\mathbf{k}}|\psi_{\mathbf{k},s} \rangle,
\end{equation}
where for simplicity the envelop function $A_{\mathbf{k}}$ is considered to be a Gaussian, $A_{\mathbf{k}} = C e^{-a^2 (\mathbf{k} - \mathbf{k_0})^{2} /2}$ for $|\mathbf{k}|\geq k_{F}^{s}$. Such a definition is valid only near the respective Fermi surfaces of the corresponding chiral sub-bands, and is similar in spirit to the definition of Landau QPWP corresponding to an SU(2) symmetric Landau Fermi liquid \cite{HK}. The basic requirements for the envelop function remain same as that have been taken in Ref. [13], i.e., $A_{\mathbf{k}}$ is a smooth function, sufficiently sharply peaked near $\mathbf{k_0}$ with spread $a^{-1} = \Delta k << k_0 - k_{F}^{s}$ and vanishes for $k<k_{F}^{s}$. This means that for all practical purposes, $k_0 \approx k_{F}^{s}$ for the c-QPWP of specific chirality `$s$'. However, on top of these the spread $\Delta k << \alpha $ which further guarantees that probability of finding a chiral quasi-particle state of specific chirality near the Fermi surface with opposite chirality is vanishingly small. This restriction is fundamentally different from the restriction (in the sense of disallow) on the inter-sub-band transition corresponding to CFL description \cite{ARM}. Furthermore, considering a similar definition of bare chiral particle wave packet one can find $\sum_{\mathbf{k}}|A_{\mathbf{k}}|^2 = 1$ by normalizing the wave packet. In this way one may think of the factor $|A_{\mathbf{k}}|^2 \simeq (2\pi)^2 \delta(\mathbf{k}-\mathbf{k_0})$, in the limit $\Delta k \rightarrow 0$ however, this is not of absolute necessity. In this article, I want to calculate the total charge and current in the c-QPWP, by supposing $s=+1$ for definiteness. However, conclusions shall remain same for $s=-1$ also. 
\paragraph*{}
The expectation values, $n(\mathbf{q}) = \langle \Psi_{\mathbf{k_0}, +} | \hat{n}(\mathbf{q})|\Psi_{\mathbf{k_0}, +} \rangle$ corresponding to the charge density, and $j(\mathbf{q}) = \langle \Psi_{\mathbf{k_0}, +} | \hat{\mathbf{j}}(\mathbf{q})|\Psi_{\mathbf{k_0}, +} \rangle$ corresponding to the current density are calculated to first order in $H_{int}$ using the non-degenerate RS perturbation theory applied to $|\psi_{\mathbf{k},+}\rangle$. The presence of Rashba SO coupling does not alter the quasi-particle properties as mentioned earlier \cite{Chen, Chesi, Aasen, Sara}, and although the states $|\mathbf{k},+\rangle$ are in continuum  the divergences  originating from this are assumed to integrable \cite{HK}. 
Within 1st order perturbation in $H_{int}$, the c-QPWP is given by,
\begin{eqnarray}\label{purt}
|\Psi_{\mathbf{k_0}, +} \rangle & =& \sum_{\mathbf{k}, |\mathbf{k}|\geq k_{F}^{+}} A_{\mathbf{k}}|\psi_{\mathbf{k},+} \rangle \nonumber \\  &=& \sum_{\mathbf{k}, |\mathbf{k}|\geq k_{F}^{+}} A_{\mathbf{k}} (|\mathbf{k},+\rangle + |\mathbf{k},+\rangle_{(1)}),
\end{eqnarray}
where $|\mathbf{k},+\rangle_{(1)} = \frac{P}{\xi_{\mathbf{k},+} - H_0} H_{int} |\mathbf{k},+\rangle$, and $H_0 |\mathbf{k},+\rangle = \xi_{\mathbf{k},+} |\mathbf{k},+\rangle$. The operator $P = ( I - |\mathbf{k},+\rangle \langle \mathbf{k},+| )$ is the projection operator which rules out the scattering of the state $|\mathbf{k},+\rangle$ by $H_{int}$ to itself, and thereby get rid of the divergences originating from $\frac{1}{\xi_{\mathbf{k},+} - H_0} $ \cite{EMS}.
\subsection{Charge density of c-QPWP}\label{2.1}
 The expectation value of $\hat{n}(\mathbf{q})$ in the state given by (\ref{purt}) consists of three terms as shown in the following equation,
\begin{eqnarray}\label{corr_charge}
& & n(\mathbf{q}) = \sum_{\mathbf{k}, |\mathbf{k}|\geq k_{F}^{+}} A^{*}_{\mathbf{k-q}}A_{\mathbf{k}} \Big[ \langle \mathbf{k-q},+|\hat{n}(\mathbf{q})|\mathbf{k},+\rangle \nonumber \\ &+& \Big( _{(1)} \langle \mathbf{k-q},+|\hat{n}(\mathbf{q})|\mathbf{k},+\rangle + \langle \mathbf{k-q},+|\hat{n}(\mathbf{q})|\mathbf{k},+\rangle_{(1)} \Big) \Big], \nonumber \\
\end{eqnarray}
where, the first term represents the charge density of a bare particle state $|\mathbf{k},+\rangle$, and the other two terms collectively represent the first order correction to the charge density in the presence of electron-electron interaction. In the following I shall consider the cases $|\mathbf{q}| = 0$, and $|\mathbf{q}| \neq 0$ separately. This is because, the $\mathbf{q} = 0$ and $\mathbf{q} \neq 0$ components of the charge density operator have very different meaning. The value of $n(\mathbf{q}=0)$, on the one hand corresponds to the average charge density of the system, while on the other hand for $|\mathbf{q}| \neq 0$, the quantity $n(\mathbf{q})$ describes the fluctuations in the charge density \cite{GDM, TH}. It is worthwhile to point out that in the analysis presented here the condition $|\mathbf{q}|<<k_{F}^{s}$ shall be considered. This ensures that the Fermi liquid picture remains valid, otherwise if a particle is scattered far away from the Fermi surface then the quasi-particle picture breaks down \cite{AGD, ARM, Vignale}.
\subsubsection{\text{For} $\mathbf{q}=0$}\label{2.1.1}
 For $|\mathbf{q}| = 0$, from (\ref{charge}) it is easy to recognize that $s' = s$ is the only possibility which gives non-zero contribution, because for $s' = -s$ the charge density operator $\hat{n}(\mathbf{q})$ vanishes identically. Therefore, in this case from the above equation and (\ref{charge}) one can find,
\begin{eqnarray}\label{chargeq0}
n(\mathbf{q}=0) &=& \sum_{\mathbf{k}, |\mathbf{k}|\geq k_{F}^{+}} \sum_{\mathbf{k}',s} |A_{\mathbf{k}}|^2  \Big[ \langle \mathbf{k},+| c_{\mathbf{k'},s}^{\dagger} c_{\mathbf{k'},s} |\mathbf{k},+\rangle \nonumber \\ & + & _{(1)} \langle \mathbf{k},+| c_{\mathbf{k'},s}^{\dagger} c_{\mathbf{k'},s} |\mathbf{k},+\rangle + h.c. \Big].
\end{eqnarray}
The first term in the square bracket leads to a value $N+1$, where $N$ represents the total number of chiral electrons present within the Fermi circles of radii $k_{F}^s$. The extra unit charge obtained above corresponds to the added particle localized within the spread $\Delta k$ of the wave packet. The second and third terms in the above equation are the first order correction to the charge density as a result of e-e interaction. These corrections can be shown to vanish for $\mathbf{q}=0$. This happens because in this case the e-e interaction does not lead to any state with an additional electron-hole pair (see Appendix \ref{app01} for detailed explanation) \cite{HK}. Therefore, $n(\mathbf{q}=0) = N + 1 $ which signifies that the total charge of the system is a constant and hence is conserved by the interaction. This is further apparent from the fact that the charge density operator commutes with the full Hamiltonian.  It is worthwhile to point out that the $\mathbf{q}=0$ component of the charge density represents the average charge density of the system because $n(\mathbf{q}=0) = \frac{1}{\Omega} \int d^2 r \; n(\mathbf{r}) = (N+1)/\Omega$. However, in this paper the volume of the system ($\Omega$) has been taken as unity and hence $n(\mathbf{q}=0) = N+1$, represents the total charge of the system. Whereas, in the case of non-zero $\mathbf{q}$ the charge density fluctuation does indeed couple to states with an additional electron-hole pair, as explained below, leading to non-zero first order correction.
\subsubsection{\text{For} $\mathbf{q}\neq 0$}\label{2.1.2}
I now calculate the charge density $n(\mathbf{q})$ for $\mathbf{q}\neq 0$. In (\ref{corr_charge}), the first term represents the charge density of a bare chiral particle wave packet. This term can be easily calculated to be $\sum_{\mathbf{k}, |\mathbf{k}|\geq k_{F}^{+}} A^{*}_{\mathbf{k-q}}A_{\mathbf{k}} \frac{1}{2}[1+  e^{-i(\theta_{\mathbf{k-q}} - \theta_{\mathbf{k}})} ]$ . By noting the fact that in the summation over $\mathbf{k}$ mentioned above $|\mathbf{k}|\geq k_{F}^{+}$ ,one can assume $\theta_{\mathbf{k}}\approx\theta_{\mathbf{k-q}}$ for $|\mathbf{q}| << k_{F}^{+}$. Therefore, the first term corresponding to (\ref{corr_charge}) turns out to be, $ \sum_{\mathbf{k}, |\mathbf{k}|\geq k_{F}^{+}} A^{*}_{\mathbf{k-q}}A_{\mathbf{k}}$. At this point I define, for notational convenience, $Q(\mathbf{q})= \sum_{\mathbf{k}, |\mathbf{k}|\geq k_{F}^{+}} Q(\mathbf{k,q}) = \sum_{\mathbf{k}, |\mathbf{k}|\geq k_{F}^{+}} A^{*}_{\mathbf{k-q}}A_{\mathbf{k}}$. This term is analytic at $\mathbf{q} = 0$, and 
\begin{equation*}
 \lim_{|\mathbf{q}| \rightarrow 0} Q(\mathbf{q}) = \sum_{\mathbf{k}} |A_{\mathbf{k}}|^2 = 1,
\end{equation*} 
where the above summation can, in principle, be performed over all $\mathbf{k}$-states since $A_{\mathbf{k}} = 0$ for all $\mathbf{k}<k_{F}^{+}$. Consequently, in this case the first term corresponding to (\ref{corr_charge}) turns out to be $Q(\mathbf{q})$.
\begin{figure*}
\centering
\includegraphics[scale=0.27]{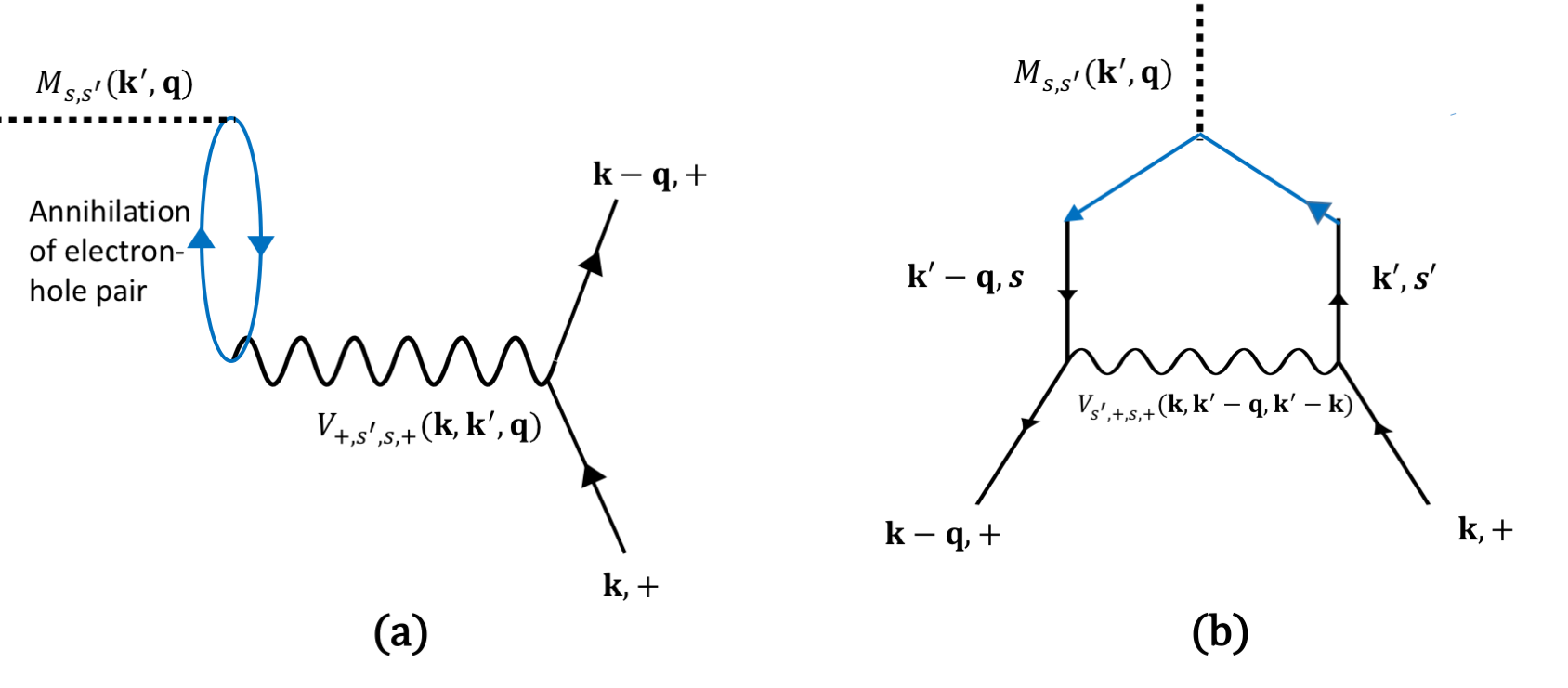}
\caption{Scattering processes leading to (\ref{nq}).  (a) electron-hole pair annihilation by $c_{\mathbf{k'-q},s}^{\dagger} c_{\mathbf{k'},s'}$, (b) annihilation of a chiral hole $\mathbf{k}_2, s_3$ and a chiral electron $\mathbf{k}_1-\mathbf{p}, s_1$ by $c_{\mathbf{k'-q},s}^{\dagger} c_{\mathbf{k'},s'}$. In figure the dashed line denoted by $M_{s, s'} (\mathbf{k', q})$ represents the form factor (or the matrix element) $\frac{1}{2}\Big[1+ s s' e^{-i(\theta_{\mathbf{k'-q}} - \theta_{\mathbf{k'}})} \Big]$ corresponding to the charge density operator (\ref{charge}). This illustrates how the charge density operator takes part in the scattering processes. Conservation of momentum and chirality index at each vertex have been taken into account. Points to note that when $M_{s, s'} (\mathbf{k', q})$ (rather a vector quantity in nature) is taken to be $\frac{1}{m}(\mathbf{k'}-\frac{\mathbf{q}}{2}) \frac{1}{2} [1+ s s' e^{-i(\theta_{\mathbf{k'-q}} - \theta_{\mathbf{k'}})}]$ the scattering processes lead to (\ref{jq+}) and when taken to be $\frac{\alpha}{2} [s' e^{i\theta_{\mathbf{k'}}} (\hat{x} - i \hat{y}) + s e^{-i\theta_{\mathbf{k'-q}}} (\hat{x} + i \hat{y})]$ these lead to (\ref{jso}).
}
\end{figure*}
\paragraph*{}
The other two terms in (\ref{corr_charge}) represent the 1st order corrections due to the electron-electron interaction as mentioned earlier. In order to compute these two terms one needs to collect  the scattering events those contribute. These are shown in FIG. 2. The first order correction $|\mathbf{k},+\rangle_{(1)}$ to the state $|\mathbf{k},+\rangle$ appearing in (\ref{corr_charge}) consists of a scattered (by the interaction) chiral electron $\mathbf{k} - \mathbf{p}, s_1$, leading to $s_4 = +$, and a chiral electron-hole pair of momentum $\mathbf{p}$. Then there can be two possibilities. In the first, the operator $c_{\mathbf{k'-q},s}^{\dagger} c_{\mathbf{k'},s'}$ can annihilate the chiral electron-hole pair as shown in the FIG. 2(a). This process leads to $\mathbf{p}= \mathbf{q},\, s_2=s', \, s_3=s, \, s_1=+,\, \text{and} \, \mathbf{k}_2 = \mathbf{k}'$. In the other process, the operator $c_{\mathbf{k'-q},s}^{\dagger} c_{\mathbf{k'},s'}$ can annihilate a chiral hole $\mathbf{k}_2, s_3$ and a chiral electron $\mathbf{k}_1 - \mathbf{p}, s_1$ as shown in FIG. 2(b). This leads to $\mathbf{p} = \mathbf{k' -k}, \, s_1 = s', \, s_3= s, \, \text{and} \, s_2 =+$.
 All the other scattering processes do not contribute because of the restrictions imposed by the projection operator $P$ as mentioned earlier. Carrying out the calculations corresponding to the scattering processes shown in FIG. 2, and adding to this the resulting expression corresponding to the first term of (\ref{corr_charge}) i.e., $Q(\mathbf{q})$,  one can find, in the limit $|\mathbf{q}|\rightarrow 0$ (see Appendix \ref{app01} for intermediate steps),
 \begin{widetext} 
\begin{equation}\label{nq+}
\lim_{|\mathbf{q}|\rightarrow 0} n(\mathbf{q}) = 1 - \sum_{\mathbf{k}, \mathbf{k'}, s} |A_{\mathbf{k}}|^{2}
 [V(\mathbf{0}) - \frac{1}{2}(1+s\cos(\theta_{\mathbf{k'}}
 -\theta_{\mathbf{k}}))V(\mathbf{k'-k})] \frac{\hat{\mathbf{k'}}\cdot \hat{\mathbf{q}} \,
  \delta(k'-k_{F}^{s})}{\frac{1}{m}(\mathbf{k'-k})\cdot \hat{\mathbf{q}} - \alpha
  (\mathbf{\hat{k}\cdot\hat{q}} - s\mathbf{\hat{k'}\cdot \hat{q}})}
\end{equation}
\end{widetext}
where $\hat{\mathbf{k'}}$ and $\hat{\mathbf{q}}$ are unit vectors.  It is worthwhile to note that for $\alpha = 0$ the chiral basis coincides with the spin basis and the expressions corresponding to (\ref{nq+}) coincides with the equations obtained in Ref. [14]. The second term of (\ref{nq+}) does not at all depend on the magnitude, $|\mathbf{q}|$ of the momentum $\mathbf{q}$, instead it depends only on the angle between $\mathbf{k}'$ and $\mathbf{q}$, and $\mathbf{k}$ and $\mathbf{q}$. Furthermore, in section \ref{3}, I have explained that in the limit $\mathbf{q} \rightarrow 0$ the second term of (15) is indeed a non-zero and positive quantity and therefore, $\lim_{|\mathbf{q}| \rightarrow 0} n(\mathbf{q}) < 1$. This signifies that in the limit $|\mathbf{q}| \rightarrow 0$ the density fluctuation does not vanish continuously and therefore, the value of $n(\mathbf{q})$ changes discontinuously from $N+1$ to a value which is less than 1 at $ \mathbf{q} = 0$. Such a discontinuity in $n(\mathbf{q})$ at $\mathbf{q}=0$ indicates a delocalization of charge as explained in section \ref{3}. This discontinuity is in addition to the one that naturally arises from the uniform charge density of the filled Fermi circle. The delocalization shall further be explored in detail in section \ref{3}, and the amount of charge delocalized will be estimated. In the following I shall calculate the current density in the c-QPWP state.
\subsection{Current density of c-QPWP}\label{2.2}
In order to compute the current density of a c-QPWP let us first note that both the kinetic part, $\hat{\mathbf{j}}_{kin}(\mathbf{q})$ and spin-orbit part, $\hat{\mathbf{j}}_{SO}(\mathbf{q})$ of the current density operator have the form, $c_{\mathbf{k'-q},s}^{\dagger} c_{\mathbf{k'},s'}$ which is same as that of the charge density operator corresponding to (\ref{charge}); they differ only by their coefficients (or matrix elements). This can be easily seen by comparing (\ref{curr_chi}) with (\ref{charge}). Therefore, both the scattering processes corresponding to FIG 2(a) and 2(b) remain same in the case of both the current densities. However, in this case the dashed line representing the matrix elements as indicated in the FIG. 2 is given by $\frac{1}{m}(\mathbf{k'}-\frac{\mathbf{q}}{2}) \frac{1}{2} [1+ s s' e^{-i(\theta_{\mathbf{k'-q}} - \theta_{\mathbf{k'}})}]$ when one consider $\hat{\mathbf{j}}_{kin}(\mathbf{q})$, and $\frac{\alpha}{2} [s' e^{i\theta_{\mathbf{k'}}} (\hat{x} - i \hat{y}) + s e^{-i\theta_{\mathbf{k'-q}}} (\hat{x} + i \hat{y})] $ for the calculation of the expectation value of $\hat{\mathbf{j}}_{SO}(\mathbf{q})$. Then by straight forward evaluation of the scattering processes corresponding to both $\hat{\mathbf{j}}_{kin}(\mathbf{q})$ and $\hat{\mathbf{j}}_{SO}(\mathbf{q})$ one can show (see Appendix \ref{app02}) in the limit $|\mathbf{q}| \rightarrow 0$ that the total current density of the c-QPWP state is given by,
\begin{widetext}
\begin{eqnarray}\label{jq}
\lim_{|\mathbf{q}|\rightarrow 0} \mathbf{j}(\mathbf{q}) &=& \lim_{|\mathbf{q}|\rightarrow 0} [\mathbf{j}_{kin}(\mathbf{q}) + \mathbf{j}_{SO}(\mathbf{q})] \nonumber \\ &=& \Bigg( \frac{\mathbf{k}_{0}}{m} + \alpha \hat{\mathbf{k}_0} \Bigg) - \sum_{\mathbf{k}, \mathbf{k'}, s} |A_{\mathbf{k}}|^{2}
 \Bigg( \frac{\mathbf{k'}}{m} + s\alpha \hat{\mathbf{k'}} \Bigg)  \frac{\hat{\mathbf{k'}}\cdot \hat{\mathbf{q}} \,
  \delta(k'-k_{F}^{s}) \Big[V(\mathbf{0}) - \frac{1}{2}(1+ s\cos(\theta_{\mathbf{k'}}
 -\theta_{\mathbf{k}}))V(\mathbf{k'-k})\Big]}{\frac{1}{m}(\mathbf{k'-k})\cdot \hat{\mathbf{q}} - \alpha
   (\mathbf{\hat{k}\cdot\hat{q}} - s\mathbf{\hat{k'}\cdot \hat{q}})}. \nonumber \\
\end{eqnarray}
\end{widetext}
Since both the kinetic part and the spin-orbit part of the current density operator commute with the full Hamiltonian the total current in the system is conserved, i.e., $\mathbf{j}(\mathbf{q = 0})=\frac{ \mathbf{k_0}}{m} + \alpha \hat{\mathbf{k}_0}$, where $\mathbf{j}(\mathbf{q = 0})$ represents the total current as $n(\mathbf{q = 0})$ represents the total charge. Therefore, the total current in the system is directed along the propagation of the c-QPWP, i.e., along $\hat{\mathbf{k}}_0$. It is worthwhile to repeat an important point that in the summations corresponding to the above equations $|\mathbf{k}|\geq k_{F}^{+}$ and $|\mathbf{k'}| = k_{F}^{s}$. Following arguments similar to those of the charge density, from (\ref{jq}) one can see that $\lim_{|\mathbf{q}| \rightarrow 0} j(\mathbf{q}) \neq  \frac{ \mathbf{k_0}}{m} + \alpha \hat{\mathbf{k}_0}$ thereby signalling in a discontinuity in the current density too at $\mathbf{q = 0}$. This leads to a delocalized current which is investigated in detail in section \ref{3}. However, it can be shown that (see Appendix \ref{app0}) the following continuity equation,
\begin{equation}\label{cont}
\frac{\partial n(\mathbf{q},t)}{\partial t}  +i \mathbf{q}\cdot \Big[\mathbf{j}_{kin}(\mathbf{q},t)+\mathbf{j}_{SO}(\mathbf{q},t)\Big] = 0
\end{equation}
is satisfied to first order in the inter-particle interaction for every $\mathbf{q}$. This implies that at each point $\mathbf{r}$ in the real space the charge-current conservation is satisfied to first order in the interaction.
\section{Delocalization of charge and current}\label{3}
Delocalization of charge can be investigated by evaluating the fluctuation in the charge density, $\Delta n(\mathbf{q})$ of the c-QPWP from its value corresponding to $|\mathbf{q}| = 0$. One can rewrite (\ref{nq}) as $n(\mathbf{q}) = Q(\mathbf{q}) + \Delta n(\mathbf{q})$ for $|\mathbf{q}| \neq 0$, and identify the fluctuation as,
\begin{eqnarray}\label{delnq}
\Delta n(\mathbf{q}) &=& n(\mathbf{q}) - Q(\mathbf{q}) \nonumber \\ &=& \sum_{\mathbf{k}} Q(\mathbf{k}, \mathbf{q}) f(\mathbf{k},\mathbf{q}),
\end{eqnarray}
where the quantity $f(\mathbf{k},\mathbf{q})$ is given by,
\begin{eqnarray}\label{f1}
& &f(\mathbf{k},\mathbf{q}) = -\sum_{\mathbf{k'}, s} \Bigg( \frac{n_0 (\mathbf{k' - q},s) -n_0 (\mathbf{k'},s)}{\frac{1}{m}(\mathbf{k'-k})\cdot \mathbf{q} - \alpha (\mathbf{\hat{k}\cdot q}-s \mathbf{\hat{k}'\cdot q})}  \nonumber \\ & &  \Big[V(\mathbf{q}) -\frac{1}{2}(1+s\cos(\theta_{\mathbf{k'}} -\theta_{\mathbf{k}}))V(\mathbf{k'-k})\Big] \Bigg).
\end{eqnarray}
This quantity represents the first order contribution to the fluctuation in the charge density from the perturbation caused by the electron electron interaction.  In (\ref{delnq}), it seems that $\Delta n(\mathbf{q}=0) =  n(\mathbf{q}=0) - 1 = N$ however, this anomaly is a result of the fact that there exists a discontinuity in the charge density which is arising from the charge density of the uniformly filled Fermi circle. Therefore, in what follows whenever I consider $\Delta n(\mathbf{q}=0)$ it is suitably redefined by subtracting $N$ from it so that at $|\mathbf{q}| =0$ the fluctuation vanishes. This section aims to show that in the limit $|\mathbf{q}| \rightarrow 0$, the fluctuation $\Delta n(\mathbf{q})$ is non-zero therefore, showing a discontinuity in the charge density at $\mathbf{q}= 0$.
\paragraph*{} 
 Owing to the sharpness of the envelop function $A_{\mathbf{k}}$ the above function (\ref{f1}) $
$ can be approximated in the small $|\mathbf{q}|$ limit as, $f(\mathbf{k_0},\mathbf{q}) = f(\theta_{\mathbf{q},\mathbf{k_0}})$, where $\theta_{\mathbf{q},\mathbf{k_0}}$ is the angle between the vector $\mathbf{q}$ and $\mathbf{k_0}$. I consider these individuals to be of the form, $\mathbf{k_0} \approx k_{F}^{+}(\cos \theta_{k_0} , \sin \theta_{k_0})$ and $\mathbf{q} = q(\cos \theta_q , \sin \theta_q)$. Therefore, in the limit $|\mathbf{q}| \rightarrow 0$ the above equation takes the form,
\begin{eqnarray}\label{f2}
& &f(\theta_{\mathbf{q},\mathbf{k_0}}) =  f^{+}(\theta_{\mathbf{q},\mathbf{k_0}}) + f^{-}(\theta_{\mathbf{q},\mathbf{k_0}}), \, \text{where,} \nonumber \\
& &f^{s}(\theta_{\mathbf{q},\mathbf{k_0}}) =
 - \sum_{\mathbf{k'}} \Bigg( \Big[V(\mathbf{0}) -\frac{1}{2}[1+ s \cos(\theta_{\mathbf{k'}} -\theta_{\mathbf{k_0}})]   \nonumber \\ & &  V(\mathbf{k'-k_0})\Big]\frac{\hat{\mathbf{k'}} \cdot \hat{\mathbf{q}}\,\delta(k'-k_{F}^{s})}{\frac{1}{m}(\mathbf{k'-k_0})\cdot \mathbf{\hat{q}} - \alpha (\mathbf{\hat{k}_0 \cdot q}-s \mathbf{\hat{k}'\cdot q})}  \Bigg),
\end{eqnarray}
and 
\begin{equation}
  \Delta n (\mathbf{q})= 
        \begin{cases}
             Q(\mathbf{q}) [ f^{+}(\theta_{\mathbf{q},\mathbf{k_0}}) + f^{-}(\theta_{\mathbf{q},\mathbf{k_0}})] & \text{ $ \mathbf{q} \neq 0$} \\
            0, & \text{ $ \mathbf{q} = 0$.} \\
        \end{cases}
\end{equation}
The function $f^{s}(\theta_{\mathbf{q},\mathbf{k_0}})$ is a sufficiently regular function, any divergences occurring due to vanishing denominator can be integrable. The finite (but small) spread of the wave packet $\Delta k = a^{-1}$ (see section \ref{2}, near (\ref{qpwp})) ensures that the denominator in the above equation does not vanish identically. In this analysis, for simplicity and ease of estimation of the amount of charge/current delocalized, I take $Q(\mathbf{q}) = e^{-q^2 a^2 /4}$, i.e., a circularly symmetric Gaussian function. However, the qualitative results do not depend on this particular choice. Using symmetry arguments one can write in two dimensions (2D),
\begin{equation}\label{expand}
 f^{s}(\theta_{\mathbf{q},\mathbf{k_0}}) = \sum_{l=0}^{\infty} f_{l}^{s} \cos(l \theta_{\mathbf{q}}) ,
\end{equation}
where $s$ denotes the chirality index. Therefore, the first order correction to the charge density in the real space is given by,
\begin{eqnarray}\label{realsp}
\Delta n(\mathbf{r})& = & \sum_{\mathbf{q}} \sum_{l=0}^{\infty} Q(\mathbf{q}) f_{l}  \cos(l \theta_{\mathbf{q}}) e^{i \mathbf{q\cdot r}},
\end{eqnarray}
where $f_{l}= (f_{l}^{+} + f_{l}^{-})$. In calculating the sum over $\mathbf{q}$'s, the standard replacement $\sum_{\mathbf{q}} \rightarrow \frac{1}{(2\pi)^2} \int d^2 q$ shall be used where the 2D volume $\Omega$ has been taken to be unity as mentioned earlier. Although I have considered the limit of small $q$, for $Q(\mathbf{q})$ sufficiently sharply peaked near $|\mathbf{q}| = 0$ the function $Q(\mathbf{q})$ vanishes everywhere except in the limit of very small values of $|\mathbf{q}| \, (\lesssim a^{-1})$. Therefore, in the above Fourier transform one can take the limit $\mathbf{q}$ integral to be from 0 to $\infty$. Following Ref. [13], I divide the real space charge density corrections in a symmetric part corresponding to $l=0$ and higher harmonic parts $l\neq 0$. Evaluating the Fourier transform it can be shown that,
\begin{eqnarray}\label{local}
\Delta n_{0}(\mathbf{r})|_{loc} &=& \frac{f_0}{2\pi a^2} e^{-\frac{r^2}{a^2}}; \, \text{with} \, f_0 = f_0^{+}+f_0^{-},
\end{eqnarray}
which represents the first order correction to localized distribution of the charge density as shown in FIG. 3(b), being finite at the origin and integrable. The localized charge density is therefore, given by $n(\mathbf{r})|_{loc} =(Q(\mathbf{r})+ \Delta n_{0}(\mathbf{r})|_{loc}) = \frac{1+ f_0}{2\pi a^2} e^{-\frac{r^2}{a^2}}$. For the charge density, the Fourier transform corresponding to the higher harmonic terms can be evaluated and it can be shown (See Appendix \ref{app1}) that the dominant behaviour at very large distance from the center of the spatial QPWP (say $r \rightarrow \infty$) is given by,
\begin{equation}\label{harm}
\Delta n_{l \neq 0}(\mathbf{r}) \approx \frac{i^l}{4\pi} \cos(l\theta) f_{l} \frac{l}{a r^2}.
\end{equation}
In the limit $r \rightarrow 0$, the charge density correction vanishes at least as $r^l$ (See Appendix \ref{app1}). The higher harmonic part of the charge density is explicitly written as, \\
\begin{equation}
       \Delta n_{l\neq 0} (\mathbf{q})= 
        \begin{cases}
            e^{-q^2 a^2 /4} f_l \cos (l\theta_{\mathbf{q}}), & \text{ $ \mathbf{q} \neq 0$} \\
            0, & \text{ $ \mathbf{q} = 0$}. \\
        \end{cases}
 \end{equation}    \\ 
The above equation along with (\ref{harm}) signify the fact that the discontinuity at $\mathbf{q} = 0$ can only provide a $r^{-2}$ tail corresponding to the higher harmonic part of the charge density. Because of this typical behavior of the higher harmonic terms they do not represent any physical distribution of charge \cite{HK}.
\paragraph*{}
From (\ref{delnq}), (\ref{expand}) and (\ref{realsp}) it is easy to see that,
\begin{equation}\label{deloc}
\Delta n_{0}(\mathbf{r}) = \frac{f_0}{2\pi a^2} e^{-\frac{r^2}{a^2}} - f_0 ,
\end{equation}
and since $\Delta n_{0}(\mathbf{q}) = 0\, \text{for} \, \mathbf{q} = 0$ it follows $\int \Delta n_{0}(\mathbf{r}) d^2 r = 0$. The first term in the above equation represents a localized charge. Therefore, the charge $-f_0 = -(f_{0}^{+}+f_{0}^{-})$ must reside at the boundary.  In Appendix \ref{app2}, I have calculated in detail the quantities $f_{0}^{+}$ and $f_{0}^{-}$, and shown that  $f_0 < 0$ for all values of electron gas parameter $r_s$. In FIG. 6, both of these are plotted as functions of the dimension less electron gas parameter $r_s$. Furthermore, it is inversely proportional to the degenerate Fermi velocity $v_0$ and therefore it depends on magnitude of the SOC.  The quantity `$-f_0$' represents a delocalized charge residing at the boundary as shown in FIG. 3(a).  However, in the case of an infinite system the boundary seems to be at the infinity which is quite unphysical. Therefore, in Appendix \ref{app3}, I have made a more quantitative estimate of the radius corresponding to the volume over which the delocalized charge is spread and shown that the delocalized charge indeed resides on the boundary of a finite volume (although this volume can be made arbitrarily large) even in an infinite system. On the other hand in the case of strong SOC the delocalized charge $f_0$ depends very strongly on the direction of propagation $\hat{\mathbf{k}}_0$ of the wave packet (see Appendix \ref{app2} for detailed calculations). Here, the quantity $f_{0}^{s}$, corresponding to the delocalized charge results from the interaction when the added bare particle of chirality `$+$' is dressed by $\frac{N}{2}$ particles present within each Fermi sphere/circle of radii $k_{F}^{+}$ and $k_{F}^{-}$.
\begin{figure*}
\centering
 \includegraphics[scale=0.45]{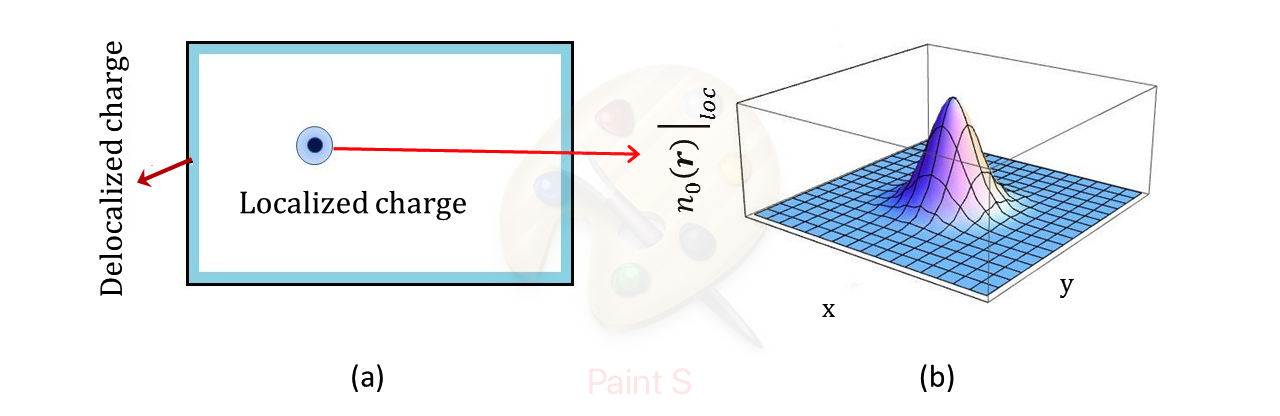}    
   \caption{(a) Schematic picture of the delocalization effect where deep blue dot corresponds to the peak of the Gaussian distribution and the blue shaded area around the dot signifies the Gaussian distribution of the localized charge within the spread of the wave packet; delocalized charge is indicated in the figure. The delocalized charge resides at the boundary of a finite system or in the case of an infinite system at  a large distance away from the centre of the c-QPWP. (b) spatial distribution of the localized charge, $n(\mathbf{r})|_{loc} = \frac{1+ f_0}{2\pi a^2} e^{-\frac{r^2}{a^2}}$.}
\end{figure*}
\paragraph*{} 
Similarly, using (\ref{jq}) in the limit $|\mathbf{q}| \rightarrow 0$,  one can define the current density fluctuations as, $\Delta \mathbf{j}(\mathbf{q}) = \mathbf{j}(\mathbf{q}) - \Big(\frac{\mathbf{k_0}}{m} + \alpha \hat{\mathbf{k}}_{0} \Big)=  Q(\mathbf{q}) \mathbf{g} (\theta_{\mathbf{q,k_0}})$, and use symmetry arguments to write,
\begin{eqnarray}\label{gq}
\mathbf{g} (\theta_{\mathbf{q,k_0}}) & = & \hat{\mathbf{k}}_0  \sum_{l=0}^{\infty} (g_l) \cos(l\theta_{\mathbf{q}})
\end{eqnarray}
where $g_l = (g_{l}^{+}+g_{l}^{-})$. Here, the vector $\mathbf{g} (\theta_{\mathbf{q,k_0}}) $ is given by, 
\begin{widetext}
\begin{equation}
\mathbf{g}(\theta_{\mathbf{q},\mathbf{k_0}}) =
 - \sum_{\mathbf{k'}, s} \Big[\frac{\mathbf{k'}}{m} + \alpha s \hat{\mathbf{k}}' \Big] \Bigg( \Bigg[V(\mathbf{0}) -\frac{1}{2}\Big[1+ s \cos(\theta_{\mathbf{k'}} -\theta_{\mathbf{k_0}})\Big]  V(\mathbf{k'-k_0})\Bigg]\frac{\hat{\mathbf{k'}} \cdot \hat{\mathbf{q}}\,\delta(k'-k_{F}^{s})}{\frac{1}{m}(\mathbf{k'-k_0})\cdot \mathbf{\hat{q}} - \alpha (\mathbf{\hat{k}_0 \cdot q}-s \mathbf{\hat{k}'\cdot q})}  \Bigg),
\end{equation}
\end{widetext}
which can be obtained by adding (\ref{jq+1}) and (\ref{jso1}) and replacing $\mathbf{k}$ by $\mathbf{k}_{0}$ in the resulting equation. In this case too the sharpness of the envelop function $A_{\mathbf{k}}$ allows the replacement. Using the above equations and repeating the steps similar to those corresponding to the charge density, the spherically symmetric part of the current density can be easily shown to be, 
\begin{eqnarray}\label{delj}
\Delta \mathbf{j}_0(\mathbf{r})& = & \frac{1}{2\pi a^2} e^{-\frac{r^2}{a^2}} g_0 \hat{\mathbf{k}}_0 - g_0 \hat{\mathbf{k}}_0
\end{eqnarray}
 where a Fourier transform of $\Delta j^{i}(\mathbf{q})$ similar to (\ref{realsp}) has been considered. Repeating the arguments corresponding to the $\Delta n_{0}(\mathbf{r})$, it is easy to see that there is a current $- g_0 \hat{\mathbf{k}}_0$ residing at the boundary. The spatial behaviours of the higher harmonic part of the current density corresponding to $l\neq 0$ are same as that of the charge density, i.e., the discontinuity at $\mathbf{q} = 0$ gives rise to a $r^{-2}$ tail in higher harmonic part of the charge density while making it vanish as $r^l$ in the limit $r\rightarrow 0$ (see Appendix \ref{app1}). This further ensures that the higher harmonic terms corresponding to the current density do not represent any net physical current available in the system. In Appendix \ref{app2}, I have shown that $g_0 > 0$ and therefore delocalized current moves in the direction opposite to the propagation of the wave packet. In the case of weak RSOC its magnitude remains same irrespective of the direction of propagation of the wave packet. On the other hand, when the RSOC is strong the magnitude of the delocalized current strongly depend on the direction of propagation of the wave packet as shown in Appendix \ref{app2}.
\section{Remarks on Experimental aspects}\label{3a} The charge and current of a bare particle are sharp quantum mechanical observables in the sense that, when a measurement is performed using a sufficiently gentle probe consisting of low frequency, long wavelength external field it produces well-defined and correct results. However, in the case of quasi-particles, properly quantifying the charge and current associated with it become exceedingly hard owing to the delocalization effect in the corresponding wave packet \cite{OH, KR}. In this case one does not know a priori, what should be the charge of the quasi-particle wave packet \cite{KR}. In the case of conventional (SU(2) invariant) Fermi liquid, owing to the spin-charge separation within the quasi-particle wave packet description the spin degree of freedom can be used to explore the delocalization effect \cite{OH}. On the contrary, in the case of Rashba spin-orbit coupled (chiral) Fermi liquid, since the spin and orbital degrees of freedom are now coupled a description of spin-charge separation is not possible. However, the presence of an extra energy scale corresponding to the RSOC provides us the required freedom. In this case, since the magnitude of the delocalized charge depends on the strength $\alpha$ of the RSOC, any measurement involving the localized charge is expected to yield results which depend on $\alpha$ since the total charge has to be equal to that of the bare particle. Although the absolute value of the magnitude of the localized charge is still ill defined because of the presence of asymptotic $r^{-2}$ tail in the charge and current, the qualitative feature corresponding to $\alpha$ dependent delocalized (and hence localized) charge should unambiguously establish the delocalization effect. It is worthwhile to point out that the charge of a chiral bare particle wave packet is 1.
 \paragraph*{}
 Recently the fundamental wave packet nature of the electron quasi-particles has been established both theoretically and experimentally in the context of reflection and transmission through barriers \cite{BOC, MAR}. Although there may be easier experimental realizations for the detection of the delocalization effect, here I shall explain the possible implications of the delocalization effect corresponding to c-QPWP within the context of this experiment. In the experiment, two indistinguishable electrons are produced on each side of a barrier by two independent source, and then they are allowed to interfere. It has been found that the probability of detecting both the particles at the same side of the barrier is non zero \cite{BOC}. The behaviour of low frequency fluctuations in the output current, which measures the probability of detecting both the electron on the same side of the barrier, has been well explained by considering the wave packet nature of the electrons \cite{BOC}. Moreover, in the above mentioned experiment the electrons are essentially (Landau) quasi-particles corresponding to SU(2) invariant Fermi liquid \cite{MAR}. Therefore, the output current mentioned above is a result of the localized charge carried by the QPWP. In the case of Rashba spin-orbit coupled electron liquid both the localized and delocalized charges of the c-QPWP depend on $\alpha$ as mentioned earlier, and strength $\alpha$ serves as an external parameter. If an experiment of type similar to the one mentioned above is carried out on 2D Rashba spin-orbit coupled electron liquid (a chiral Fermi liquid) where the strength of RSOC, $\alpha$ is tunable externally, then an $\alpha$ dependent output current would unambiguously establish the delocalization effect. Such a chiral Fermi liquid is commonly realized in 2D semiconductor heterostructure.
\section{Conclusions and discussions}\label{4}
In this paper, I have investigated the internal structure of a chiral quasi-particle wave packet (c-QPWP) of average momentum $\mathbf{k_0}$, and the delocalization effect caused by the inter-particle interactions. The c-QPWP is defined as the conventional superposition of chiral quasi-particle states. The validity of the definition is limited near the respective chiral Fermi surfaces. The definition of c-QPWP adopted here is very similar in spirit to the Landau QPWP \citep{HK}. It is found that the interaction between the chiral electrons indeed expels some charge and current to the boundary of the system. The internal structure of the c-QPWP has the following properties. The charge/particle density consists of three parts: a spherically symmetric part indicating a localized charge corresponding to the QPWP; this part when integrated gives the value $e^{'}$ of its charge less than the bare charge of chiral quasi-particle. The second part is a higher harmonic part which vanishes at the origin and behaves as $r^{-2}$ far away from the origin. Since the higher harmonic part corresponding to the charge density dies out far away from the centre of the QPWP and at the same time vanishes at the centre/origin, this part does not represent any net charge available in the system. Finally, the remaining part represents the effect of interaction and signifies a charge $1-e^{'}$ which resides at the boundary of the system. Therefore, total charge in the system is 1 (modulo $N$, which is the total charge present within the Fermi surface expressed in the units of electronic charge $e$). However, the amount of charge expelled to the boundary depends on the strength of the SO coupling and turns out to be inversely proportional to the degenerate Fermi velocity $v_0$ when the RSOC is weak. At this point it is worthwhile to point out that the effect of electron-electron interaction corresponding to the QPWP is the delocalization of charge, which turns out to be quite general be it a Landau-QPWP or c-QPWP. Higher order contributions corresponding to the perturbation is not expected to alter the qualitative results corresponding to the SO strength dependent delocalization of charge \cite{Stamp}. Similar decomposition of current density has been found and the spatial behaviour of remains same. Fractions of current are expelled to the boundary although the bare chiral particle wave packet is localized. Interestingly enough, both the charge and current corresponding to the c-QPWP contains effects from both the Rashba sub-bands. Although I have started with a quasi-particle of specific chirality, the magnitude of delocalization turns out to be a sum total of the effect of e-e interaction both within the sub-band and inter sub-band. This additive nature of the contributions (within the first order perturbation in interaction) from both the sub-bands further seems to indicate that the delocalization of charge and current in a two component Fermi liquid should be similar. Therefore the internal structure of the QPWP corresponding to the two-component Fermi liquid is expected to be the same \cite{OA}
\paragraph*{}
Furthermore, when the strength of RSOC becomes strong, i.e., $m\alpha^2 /2 >> \mu$, the Fermi velocities corresponding to two sub-bands no longer remain degenerate. Within first order perturbation the difference between the Fermi velocities is given by $v_+ -v_- = \frac{1}{\pi} \ln(\frac{k_{F}^{+}}{k_{F}^{-}})$, where the Fermi momenta are now renormalized one \cite{ARM}. However, this does not alter the qualitative nature of the internal structure of c-QPWP, and the delocalization effect as well. Instead, this makes the magnitudes of both the delocalized charge and the current to depend strongly on the direction of propagation of the wave packet. Moreover, the magnitude of both the delocalized charge and current depend on the strength on the RSOC quite non-trivially as shown in (\ref{f02}) and (\ref{g02}).
\paragraph*{}
 A similar analysis of the spin-density and the spin current of the c-QPWP would be more interesting, and important too from the experimental point of view. This shall be taken up in the future. Since spin is not a conserved quantity in the presence of spin-orbit coupling the conventional method of continuity equation fails in determining the form of the spin current \cite{Ras}. The non-uniqueness of the definition of equilibrium spin-current in the presence of spin-orbit coupling makes the analysis more challenging \cite{Ras, Shi, Bray}.
 \paragraph*{}
 I now emphasize the special role played by the spin-orbit coupling (SOC) in the delocalization effect. The SOC provides as an extra energy scale available in the system apart from the Fermi energy and energy corresponding to electron-electron interaction. Furthermore, the quasi-particle properties largely remain unaltered even though the SOC is present \cite{Chen, Chesi, Aasen, Sara}. The delocalization of charge and current of a c-QPWP obtained here are caused by the interaction between the electrons. However, it is only because of the presence of SOC that the delocalized charge and current turn out to be a function of the strength $\alpha$ of RSOC. This stems from the fact that because of the SOC the orbital and spin degrees of freedom of the electron are now locked and the Fermi surface become spin split. The system develops two concentric Fermi circles with their radii as a function of the strength of SOC. The chiral electrons present inside both the Fermi circles interact with the added bare chiral electron outside the Fermi circle, thereby making the delocalized charge and current to depend on $\alpha$. On the contrary, the absence of SOC and consequently the presence of a single Fermi surface forbids any such feature to appear in the case of SU(2) invariant Landau Fermi liquid. This is a no ordinary effect caused by the SOC in view of the fact that in most of the experiments with 2D semiconductor heterostructure the strength of the SOC is externally tunable. Therefore, by tuning the strength $\alpha$ of the SOC, if one finds an $\alpha$ dependent output in an experiment performed on 2D chiral Fermi liquid it would unambiguously establish the delocalization effect. Experiments similar to that reported in Ref. [15] (and theoretically analysed in Ref. [16]) may be performed to see if and how the measured output currents depend on the strength of the RSOC.
 \paragraph*{}
 With linear Dresselhaus spin-orbit coupling (DSOC) instead of RSOC the results remain the same even quantitatively \cite{Dres}. This is because, in case of DSOC the phase $\theta_{\mathbf{k}}$ corresponding to the chiral bare particle states differs by $\pi /2$ from the case of RSOC and phases cancel out in expectation values of every density operators mentioned above. However, if both DSOC and RSOC are present the situations should be a topic of further investigations which could be a natural extension too.
\section*{Acknowledgements}
The author acknowledges Olle Heinonen, Ranjan Chaudhury, and Paramita Dutta for useful comments and suggestions. The author further acknowledges Mukunda P. Das, and Saptarshi Mandal for useful discussions. The author would like to express his appreciation for the valuable suggestions and criticisms of the referee in preparing the revised manuscript.
\begin{widetext}
\begin{appendix}
\section{Detailed calculations of charge and current densities}\label{app00}
\subsection{Calculation for the charge density}\label{app01}
\textit{For $q=0$}: In the following I shall describe the steps to show that in the caseof $q=0$ the first order correction indeed vanishes. The physical picture behind this is the fact that although the interaction scatters the state $|\mathbf{k},+\rangle$ to states with an additional electron-hole pair the average density operator does not couple to any such state. The net effect is the fact that the interaction effect can't provide any state with an additional electron-hole pair. To see this let us first note that the projection operator $P$ allows all the states scattered by $H_{int}$, except $|\mathbf{k},+\rangle$. Suppose, $H_{int}$ scatters the state $|\mathbf{k},+\rangle$ to states $|\mathbf{k}'',s''\rangle$ with amplitude $B_{\mathbf{k}'',s''}$ (with restriction $\mathbf{k}''\neq \mathbf{k}$, and $s''\neq +$). Then the terms corresponding to the first order correction become,\begin{equation*}
\sum_{\mathbf{k}} \sum_{\mathbf{k',s}} \Big[ \Big( \sum_{\substack{\mathbf{k''(\neq\mathbf{k})}, \\ s''(\neq +)} } B_{\mathbf{k}'',s''} \langle c_{\mathbf{k},+} c_{\mathbf{k'},s}^{\dagger} c_{\mathbf{k'},s} c_{\mathbf{k''},s''}^{\dagger} \rangle \Big) +h.c. \Big],
\end{equation*} 
where the expectation value corresponding to the above equation is taken in the state $|F\rangle $. Using Wicks theorem one can easily show that the above terms vanish because of the restriction $\mathbf{k}''\neq \mathbf{k}$ and $s''\neq +$. These lead to the result that $n(\mathbf{q}=0) = N+1$. Furthermore, $n(\mathbf{q}) = \frac{1}{\Omega} \int d^2 r  e^{- i \mathbf{q\cdot r}} n(\mathbf{r})$ is the Fourier transform of the charge density operator corresponding to the real space. With $\mathbf{q} = 0$ and $\Omega=1$, one finds $n(\mathbf{q}=0) =  \int d^2 r\; n(\mathbf{r})$, i.e. the total charge. Whereas, in the case of non-zero $\mathbf{q}$ the charge density operator indeed represent density fluctuation which can then couple to states with an additional electron-hole pair.
\paragraph*{}
\textit{For $q\neq 0$}: I now explain the intermediate steps to obtain (\ref{nq+}) corresponding to section \ref{2.1.2}. In this case one can carry out the calculations corresponding to the scattering processes shown in FIG. 2 to obtain, 
\begin{eqnarray}\label{nq}
 n(\mathbf{q}) =  Q(\mathbf{q}) - \sum_{\mathbf{k}, \mathbf{k'}, s, s'} Q(\mathbf{k,\, q}) \frac{1}{2} \Big[1+ s s' e^{-i(\theta_{\mathbf{k'-q}} - \theta_{\mathbf{k'}})} \Big]  & & \frac{n_0 (\mathbf{k' - q},s') -n_0 (\mathbf{k'},s')}{\frac{1}{m}(\mathbf{k'-k})\cdot \mathbf{q} - \alpha (\mathbf{\hat{k}\cdot q}-s' \mathbf{\hat{k'}\cdot q})} \times  \nonumber \\ & & \Big[V_{+,s',s,+}(\mathbf{k,k'}, \mathbf{q}) - V_{s',+,s,+}(\mathbf{k, k'-q},\mathbf{k'-k})\Big], 
\end{eqnarray}
where the sum over $\mathbf{k}$ is restricted for values $|\mathbf{k}|\geq k_{F}^{+}$. In obtaining the above expression any occurrence of $(\frac{q}{k_{F}^{+}})^{2}$ has been neglected by assuming $q<<k_{F}^{+}$. Here $n_0 (\mathbf{k},s)$ denotes the expectation values in the non-interacting ground state, and the term $[n_0 (\mathbf{k' - q},s') -n_0 (\mathbf{k'},s')]$ appearing in the numerator of the above equation determines the phase space available for the scattering events to occur. In the limit of small $|\mathbf{q}|$ this is proportional to $\mathbf{q}$. This can be seen when the following simplifications are performed in the small $|\mathbf{q}|(=q)$ limit,
\begin{eqnarray}\label{phase}
 n_{0}(\mathbf{k'-q}, s) - n_{0}(\mathbf{k'}, s) = -(\xi_{\mathbf{k'-q}, s} - \xi_{\mathbf{k'}, s}) \Bigg(\frac{\partial \xi_{\mathbf{k'}, s}}{\partial k'} \Bigg) ^{-1} \delta(k'-k_{F}^{s}) &=& \frac{(1/m + s\alpha /k')\mathbf{k'\cdot q}}{(1/m + s\alpha /k')k'} \delta(k'-k_{F}^{s}) \nonumber \\
&=& \mathbf{\hat{k'}}\cdot\mathbf{q} \delta(k'-k_{F}^{s}),
\end{eqnarray}
where terms of $O(q^2)$ have been neglected. Therefore, because of the appearance of $\delta(k'-k_{F}^{s})$ the magnitude of $\mathbf{k'}$ in the summation turns out to be fixed at $k_{F}^{s}$ and  for small values of $|\mathbf{q}|$, i.e., $q<<k_{F}^{+}$ it can safely be assumed that $\theta_{\mathbf{k'}}\approx\theta_{\mathbf{k'-q}}$.  With  this condition, the term $\frac{1}{2} [1+ s s' e^{-i(\theta_{\mathbf{k'-q}} - \theta_{\mathbf{k'}})}]$ appearing in (\ref{charge}) and (\ref{curr_chi}) becomes $\frac{1}{2} [1+ s s']$, which vanishes when $s'=-s$ thereby allowing only $s'=s$. This signifies that the inter-subband scattering processes do not occur. In the limit $|\mathbf{q}|(=q) \rightarrow 0$, the equation (\ref{nq}) takes the form corresponding to (\ref{nq+}) when (\ref{phase}) is incorporated in (\ref{nq}).
\subsection{Calculations for current density}\label{app02}
By straight forward evaluation of the scattering processes corresponding to $\hat{\mathbf{j}}_{kin}(\mathbf{q})$ as given in (\ref{curr_chi}), it can be shown that in the limit of very small $|\mathbf{q}|$, 
\begin{eqnarray}\label{jq+}
\mathbf{j}_{kin}(\mathbf{q})  = \sum_{\mathbf{k}} Q(\mathbf{k,q})\Bigg( \frac{1}{m}\left(\mathbf{k} - \mathbf{q}/2 \right) &-& \sum_{\mathbf{k'},s} \frac{1}{m}\left(\mathbf{k'} - \mathbf{q}/2 \right) \frac{n_0 (\mathbf{k' - q},s) -n_0 (\mathbf{k'},s)}{\frac{1}{m}(\mathbf{k'-k})\cdot \hat{\mathbf{q}} - \alpha
  (\mathbf{\hat{k}\cdot\hat{q}} - s\mathbf{\hat{k'}\cdot \hat{q}})}\nonumber \\ & &\Big[V(\mathbf{q}) - \frac{1}{2}(1+s\cos(\theta_{\mathbf{k'}}-\theta_{\mathbf{k}}))V(\mathbf{k'-k})\Big] \Bigg).
\end{eqnarray}
In the limit $|\mathbf{q}| \rightarrow 0$ the above equation takes the form,
\begin{eqnarray}\label{jq+1}
\lim_{|\mathbf{q}|\rightarrow 0} \mathbf{j}_{kin}(\mathbf{q}) = \frac{\mathbf{k}_{0}}{m} - \sum_{\mathbf{k}, \mathbf{k'}, s} |A_{\mathbf{k}}|^{2}
 \frac{\mathbf{k'}}{m} \Big[V(\mathbf{0}) - \frac{1}{2}(1+ s\cos(\theta_{\mathbf{k'}}
 -\theta_{\mathbf{k}}))V(\mathbf{k'-k})\Big] \frac{\hat{\mathbf{k'}}\cdot \hat{\mathbf{q}} \,
  \delta(k'-k_{F}^{s})}{\frac{1}{m}(\mathbf{k'-k})\cdot \hat{\mathbf{q}} - \alpha
   (\mathbf{\hat{k}\cdot\hat{q}} - s\mathbf{\hat{k'}\cdot \hat{q}})}.\nonumber \\
\end{eqnarray}
Similarly the expectation value of the spin-orbit part of the current density given in (\ref{curr_chi}), i.e., $\mathbf{j}_{SO}(\mathbf{q})=\langle \Psi_{\mathbf{k_0}, +} |\hat{\mathbf{j}}_{SO}(\mathbf{q})|\Psi_{\mathbf{k_0}, +} \rangle$ in the chiral-QPWP state can be obtained to be,
\begin{eqnarray}\label{jso}
\mathbf{j}_{SO}(\mathbf{q}) = \sum_{\mathbf{k}} Q(\mathbf{k,q}) \Bigg( \alpha \hat{\mathbf{k}} - \sum_{\mathbf{k'},s} s\alpha \hat{\mathbf{k'}} \frac{n_0 (\mathbf{k' - q},s) -n_0 (\mathbf{k'},s)}{ \frac{1}{m}(\mathbf{k'-k})\cdot \hat{\mathbf{q}} - \alpha
   (\mathbf{\hat{k}\cdot\hat{q}} - s\mathbf{\hat{k'}\cdot \hat{q}})} \Big[V(\mathbf{q}) - \frac{1}{2}(1+s\cos(\theta_{\mathbf{k'}}-\theta_{\mathbf{k}}))V(\mathbf{k'-k})\Big] \Bigg), \nonumber \\
\end{eqnarray}
and in the limit $|\mathbf{q}| \rightarrow 0$ the above equation takes the form,
\begin{eqnarray}\label{jso1}
\lim_{|\mathbf{q}| \rightarrow 0}  \mathbf{j}_{SO}(\mathbf{q}) = \alpha \hat{\mathbf{k}_0} - \sum_{\mathbf{k, k'},s} |A_{\mathbf{k}}|^{2}s\alpha \hat{\mathbf{k'}} \frac{\hat{\mathbf{k'}}\cdot \hat{\mathbf{q}} \, \delta(k'-k_{F}^{s})}{ \frac{1}{m} (\mathbf{k'-k})\cdot \mathbf{q} - \alpha (\mathbf{\hat{k}\cdot\hat{q}} - s\mathbf{\hat{k'}\cdot \hat{q}})}\Big[V(\mathbf{0}) - \frac{1}{2}(1+s\cos(\theta_{\mathbf{k'}}-\theta_{\mathbf{k}}))V(\mathbf{k'-k})\Big]. \nonumber \\
\end{eqnarray}
Then adding (\ref{jq+1}) and (\ref{jso1}) one can arrive at the equation (\ref{jq}).
\section{Continuity equation}\label{app0}
The time evolution of the operators are considered in the interaction representation and the time dependent charge density operator is given by,
\begin{equation}\label{nqt}
\hat{n}(\mathbf{q}, t) = e^{i \mathcal{H}_0 t} \hat{n}(\mathbf{q}) e^{-i \mathcal{H}_0 t},
\end{equation} 
where $\hat{n}(\mathbf{q})$ is given by (\ref{charge}). A similar expression corresponding to $ \hat{\mathbf{j}}_{kin}(\mathbf{q},t)$ and $ \hat{\mathbf{j}}_{SO}(\mathbf{q},t)$ shall also be considered in the interaction representation in order to derive the continuity equation. The time dependent chiral-QPWP state is given by,
\begin{equation}\label{tpacket}
|\Psi_{\mathbf{k}_0, +} (t)\rangle = \left( 1- i \int_{0}^{t} dt' e^{i \mathcal{H}_0 t'} \mathcal{H}_{int} e^{-i \mathcal{H}_0 t'} \right) |\Psi_{\mathbf{k}_0, +} \rangle ,
\end{equation}
where only the first order contribution from the interaction has been taken into account. In order to arrive at the continuity equation I first derive $n(\mathbf{q},t) = \langle \Psi_{\mathbf{k}_0, +} (t)| \hat{n}(\mathbf{q}, t) |\Psi_{\mathbf{k}_0, +} (t) \rangle $, for $\mathbf{q} \neq 0$, within the first order perturbation theory. Using (\ref{qpwp}), (\ref{nqt}) and (\ref{tpacket}), the expression for the time dependent charge density can be calculated to be,
\begin{eqnarray}\label{nqt1}
& & n(\mathbf{q},t) =\langle \Psi_{\mathbf{k}_0, +} (t)| \hat{n}(\mathbf{q},t) |\Psi_{\mathbf{k}_0, +} (t)\rangle \nonumber \\
&=& \sum_{\mathbf{k}} Q(\mathbf{k,q}) e^{i(\epsilon_{\mathbf{k-q},+}-\epsilon_{\mathbf{k},+})t} \left[ 1- \sum_{\mathbf{k'},s} \frac{n_0 (\mathbf{k' - q},s) -n_0 (\mathbf{k'},s)}{\frac{1}{m}(\mathbf{k'-k})\cdot \mathbf{q} - \alpha (\mathbf{\hat{k}\cdot q} -s \mathbf{\hat{k}'\cdot q})}  \Big[V(\mathbf{q}) - \frac{1}{2}(1+s \cos(\theta_{\mathbf{k'}}-\theta_{\mathbf{k}}))V(\mathbf{k'-k})\Big] \right]
 \nonumber \\
& -i & \sum_{\mathbf{k,k'}} A_{\mathbf{k-q}}^{*}A_{\mathbf{k}} \, \langle \mathbf{k-q},+ | e^{i \mathcal{H}_0 t} \hat{n}(\mathbf{q}) e^{-i \mathcal{H}_0 t} \left( \int_{0}^{t} dt' e^{i \mathcal{H}_0 t'} \mathcal{H}_{int} e^{-i \mathcal{H}_0 t'} \right) |\mathbf{k},+ \rangle
 \nonumber \\
  & +i & \sum_{\mathbf{k,k'}} A_{\mathbf{k-q}}^{*}A_{\mathbf{k}} \, \langle \mathbf{k-q},+ | \left( \int_{0}^{t} dt' e^{i \mathcal{H}_0 t'} \mathcal{H}_{int} e^{-i \mathcal{H}_0 t'} \right) e^{i \mathcal{H}_0 t} \hat{n}(\mathbf{q}) e^{-i \mathcal{H}_0 t}  |\mathbf{k},+ \rangle .
\end{eqnarray}
In the last two terms corresponding to the above equation, only the self energy part corresponding to the FIG. 4 survive and all the other scattering events corresponding to FIG. 5 get canceled out. Carrying out the time integration it can be found that,
\begin{eqnarray}
& & n(\mathbf{q},t) \nonumber \\
&=& \sum_{\mathbf{k}} Q(\mathbf{k,q}) e^{i(\xi_{\mathbf{k-q},+}-\xi_{\mathbf{k},+})t} \Bigg( 1- it \Big[ \Sigma(\mathbf{k}) - \Sigma(\mathbf{k-q}) \Big]
\nonumber \\ &-& \sum_{\mathbf{k'},s} \frac{n_0 (\mathbf{k' - q},s) -n_0 (\mathbf{k'},s)}{\frac{1}{m}(\mathbf{k'-k})\cdot \mathbf{q} - \alpha (\mathbf{\hat{k}\cdot q} -s \mathbf{\hat{k}'\cdot q})}\Big[V(\mathbf{q}) - \frac{1}{2}(1+s\cos(\theta_{\mathbf{k'}}-\theta_{\mathbf{k}}))V(\mathbf{k'-k})\Big] \Bigg),
\end{eqnarray}
where $\Sigma(\mathbf{k}) = \sum_{\mathbf{k'},s} [V(0) - \frac{1}{2} \lbrace 1+s \cos(\theta_{\mathbf{k'}}- \theta_{\mathbf{k}}) V(\mathbf{k' - k})\rbrace] n_0 (\mathbf{k'},s)$ is the first order self energy (corresponding to FIG. 4). Similarly, it is easy to compute the kinetic part of the time dependent current density which is given by,
\begin{eqnarray}\label{jkinqt}
& & \mathbf{j}_{kin}(\mathbf{q},t)  =\langle \Psi_{\mathbf{k}_0, +} (t)| \hat{\mathbf{j}}_{kin}(\mathbf{q},t)|\Psi_{\mathbf{k}_0, +} (t)\rangle \nonumber \\
&=& \sum_{\mathbf{k}} Q(\mathbf{k,q}) e^{i(\xi_{\mathbf{k-q},+}-\xi_{\mathbf{k},+})t} \Bigg( \frac{1}{m}\left(\mathbf{k} - \mathbf{q}/2 \right)- it \frac{1}{m} \left(\mathbf{k} - \mathbf{q}/2 \right) \Big[ \Sigma(\mathbf{k}) - \Sigma(\mathbf{k-q}) \Big]
\nonumber \\ &-& \sum_{\mathbf{k'},s} \frac{1}{m}\left(\mathbf{k'} - \mathbf{q}/2 \right) \frac{n_0 (\mathbf{k' - q},s) -n_0 (\mathbf{k'},s)}{\frac{1}{m}(\mathbf{k'-k})\cdot \mathbf{q} - \alpha (\mathbf{\hat{k}\cdot q} -s \mathbf{\hat{k}'\cdot q})}\Big[V(\mathbf{q}) - \frac{1}{2}(1+s\cos(\theta_{\mathbf{k'}}-\theta_{\mathbf{k}}))V(\mathbf{k'-k})\Big] \Bigg).
\end{eqnarray}
Furthermore, by expressing (\ref{curr_chi}) in the interaction representation one can obtain the corresponding time dependent spin-orbit part of the current density operator, and using (\ref{tpacket}) it can be easily shown that,
\begin{eqnarray}\label{jsoqt}
& & \mathbf{j}_{SO}(\mathbf{q},t) =\langle \Psi_{\mathbf{k}_0, +} (t)|  \hat{\mathbf{j}}_{SO}(\mathbf{q},t) |\Psi_{\mathbf{k}_0, +} (t)\rangle  \nonumber \\
&=& \sum_{\mathbf{k}} Q(\mathbf{k,q}) e^{i(\xi_{\mathbf{k-q},+}-\xi_{\mathbf{k},+})t} \Bigg( \alpha \hat{\mathbf{k}}- it \alpha \hat{\mathbf{k}} \Big[ \Sigma(\mathbf{k}) - \Sigma(\mathbf{k-q}) \Big]
\nonumber \\ &-& \sum_{\mathbf{k'},s} s\alpha \hat{\mathbf{k'}} \frac{n_0 (\mathbf{k' - q},s) -n_0 (\mathbf{k'},s)}{ \frac{1}{m} (\mathbf{k'-k})\cdot \mathbf{q} - \alpha (\mathbf{\hat{k}\cdot q} -s \mathbf{\hat{k}'\cdot q})}\Big[V(\mathbf{q}) - \frac{1}{2}(1+s\cos(\theta_{\mathbf{k'}}-\theta_{\mathbf{k}}))V(\mathbf{k'-k})\Big] \Bigg).
\end{eqnarray}
\begin{figure}[h]
\centering
\begin{minipage}[b]{0.4\textwidth}
\centering
\includegraphics[scale=0.43]{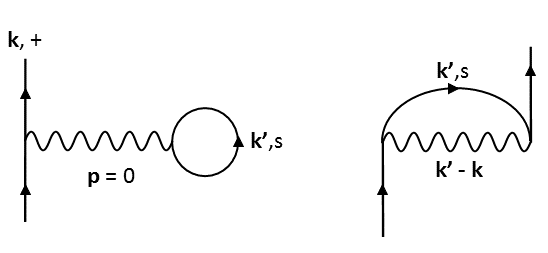}
\caption{Self-energy diagrams.}
\end{minipage}
\hspace{0.1\textwidth} 
\begin{minipage}[b]{0.4\textwidth}
\centering
\includegraphics[scale=0.40]{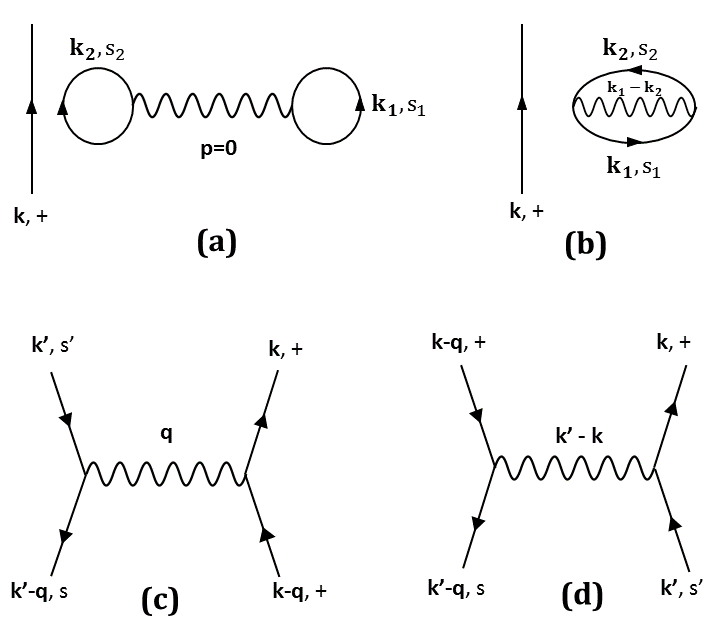}
\caption{Other scattering processes.}
\end{minipage}
\end{figure}
Furthermore, it can be easily shown that, $\Sigma(\mathbf{k}) - \Sigma(\mathbf{k-q}) = \sum_{\mathbf{k'},s} [n_{0}(\mathbf{k' - q}, s)-n_{0}(\mathbf{k'}, s)]\frac{1}{2}[1+s\cos(\theta_{\mathbf{k'}}- \theta_{\mathbf{k}})]V(\mathbf{k' -k})$, when we neglect the difference between $\theta_{\mathbf{k}}$ and $\theta_{\mathbf{k-q}}$ for small values of $|\mathbf{q}|$. This is going to be used in the continuity equation below. For $\mathbf{q} \neq 0$ one can show that the continuity equation is indeed satisfied in the following way,
\begin{eqnarray}
& &\frac{\partial n(\mathbf{q},t)}{\partial t} + i \, \mathbf{q}\cdot [j_{kin}(\mathbf{q},t) + j_{SO}(\mathbf{q},t)] = \nonumber \\ &i& \sum_{\mathbf{k}} Q(\mathbf{k,q}) e^{i(\xi_{\mathbf{k-q},+} - \xi_{\mathbf{k},+})} \Bigg[ -\Big(\frac{1}{m}(\mathbf{k} - \frac{\mathbf{q}}{2})\cdot \mathbf{q} + \alpha \hat{\mathbf{k}}\cdot \mathbf{q}\Big) \Big\lbrace 1 - it [\Sigma(\mathbf{k}) - \Sigma(\mathbf{k-q})]  \nonumber \\ &-& \sum_{\mathbf{k'},s} \frac{n_0 (\mathbf{k' - q},s) -n_0 (\mathbf{k'},s)}{\frac{1}{m}(\mathbf{k'-k})\cdot \mathbf{q} - \alpha (\mathbf{\hat{k}\cdot q} -s \mathbf{\hat{k}'\cdot q})}[V(\mathbf{q}) - \frac{1}{2}(1+s\cos(\theta_{\mathbf{k'}}-\theta_{\mathbf{k}}))V(\mathbf{k'-k})] \Big\rbrace - [\Sigma(\mathbf{k}) - \Sigma(\mathbf{k-q})] \Bigg] \nonumber \\ &+& i \sum_{\mathbf{k}} Q(\mathbf{k,q}) e^{i(\xi_{\mathbf{k-q},+} - \xi_{\mathbf{k},+})} \Bigg[ \Big(\frac{1}{m}(\mathbf{k} - \frac{\mathbf{q}}{2})\cdot \mathbf{q} + \alpha \hat{\mathbf{k}}\cdot \mathbf{q}\Big) \Big\lbrace 1 - it [\Sigma(\mathbf{k}) - \Sigma(\mathbf{k-q})] \Big\rbrace \nonumber \\ &-& \sum_{\mathbf{k'},s} \Big(\frac{1}{m}(\mathbf{k'} - \frac{\mathbf{q}}{2})\cdot \mathbf{q} + s \alpha \hat{\mathbf{k'}}\cdot \mathbf{q}\Big) \frac{n_0 (\mathbf{k' - q},s) -n_0 (\mathbf{k'},s)}{\frac{1}{m}(\mathbf{k'-k})\cdot \mathbf{q} - \alpha (\mathbf{\hat{k}\cdot q} -s \mathbf{\hat{k}'\cdot q})}[V(\mathbf{q}) - \frac{1}{2}(1+s\cos(\theta_{\mathbf{k'}}-\theta_{\mathbf{k}}))V(\mathbf{k'-k})] \Bigg] \nonumber \\ &=& i \sum_{\mathbf{k}} Q(\mathbf{k,q}) e^{i(\xi_{\mathbf{k-q},+} - \xi_{\mathbf{k},+})}  \Bigg[ \sum_{\mathbf{k'},s} (-1) \Big[\frac{1}{m}(\mathbf{k'-k})\cdot \mathbf{q} - \alpha (\mathbf{\hat{k}\cdot q}-s \mathbf{\hat{k'}\cdot q})\Big] \frac{n_0 (\mathbf{k' - q},s) -n_0 (\mathbf{k'},s)}{\frac{1}{m}(\mathbf{k'-k})\cdot \mathbf{q} - \alpha (\mathbf{\hat{k}\cdot q} -s \mathbf{\hat{k}'\cdot q})}\Big[V(\mathbf{q}) \nonumber \\ &-& \frac{1}{2}(1+s\cos(\theta_{\mathbf{k'}}-\theta_{\mathbf{k}}))V(\mathbf{k'-k})\Big] - \Big[\Sigma(\mathbf{k}) - \Sigma(\mathbf{k-q})\Big] \Bigg] \nonumber \\ &=&
 i \sum_{\mathbf{k}} Q(\mathbf{k,q}) e^{i(\xi_{\mathbf{k-q},+} - \xi_{\mathbf{k},+})}  \Bigg[ \sum_{\mathbf{k'},s} (-1) [n_0 (\mathbf{k' - q},s) -n_0 (\mathbf{k'},s)]\Big[V(\mathbf{q}) - \frac{1}{2}(1+s\cos(\theta_{\mathbf{k'}}-\theta_{\mathbf{k}}))V(\mathbf{k'-k})\Big] \nonumber \\ &-& \Big[\Sigma(\mathbf{k}) - \Sigma(\mathbf{k-q})\Big] \Bigg] = 0, 
\end{eqnarray}
since $\sum_{\mathbf{k'},s} \Big[n_{0}(\mathbf{k' - q}, s)-n_{0}(\mathbf{k'}, s)\Big] V(\mathbf{q}) = 0$. On the other hand, for $\mathbf{q} = 0$ the continuity equation is trivially satisfied. Therefore, to the first order in the interaction the continuity equation is satisfied at each point $\mathbf{r}$.
\section{Calculation of asymptotic behaviour of higher harmonic terms}\label{app1}
From (\ref{realsp}), it is easy to see that,
\begin{eqnarray}
\Delta n_0 (\mathbf{r}) &=& \frac{f_0}{4\pi} \int dq \, q \, J_{0}(qr) e^{-q^2 a^2 /4},
\end{eqnarray}
which when evaluated lead to (\ref{local}), where the Fourier-Bessel expansion, $e^{i \mathbf{q\cdot r}} =e^{i qr\cos(\theta_{\mathbf{q}}- \theta)} = \sum_{n=0}^{\infty} i^n J_n (qr) \cos(l\theta_{\mathbf{q}}-l\theta)$ has been used, and $J_n (qr)$ is the Bessel function of first kind. Here the angle $\theta$ corresponds to $\mathbf{r} = r(\cos\theta, \, \sin\theta)$. Similarly for $l\neq 0$,
\begin{equation}
\Delta n_{l} (\mathbf{r}) =  i^l \frac{f_l}{4\pi} \cos(l\theta) \frac{r}{a^3} \sqrt{\pi} e^{-r^2 / 2a^2}\left[I_{\frac{l-1}{2}}\left(\frac{r^2}{2a^2}\right) - I_{\frac{l+1}{2}}\left(\frac{r^2}{2a^2}\right) \right],
\end{equation}
where $I_{n}(z)$ is the modified Bessel function of first kind \cite{AS} and in the Fourier transform corresponding to (\ref{realsp}) only even integer values of $l$ survive. In the limit $z\rightarrow 0$, $I_{n}(z) \approx (\frac{z}{2})^{n} [\Gamma(n+1)]^{-1}$ for all $n>0$ which is satisfied for all $l\neq 0$ in the above equation \cite{AS}. Therefore, $\lim_{r \rightarrow 0} \Delta n_{l}(\mathbf{r}) \sim  i^l \frac{f_l}{4\pi} \cos(l\theta) \frac{r}{a^2} 2\sqrt{\pi} e^{-r^2 / 2a^2} \frac{r^l}{(2a)^{l}}$, i.e., the first order correction to the charge density is regular at the origin and vanishes as $r^l$. On the other hand,  one can use the asymptotic expression (in the limit of very large $z$) for $I_{n}(z)$ \cite{AS}, and from the above equation it can be easily shown that the dominant behaviour for $r >> a$ is, $\Delta n_{l}(\mathbf{r}) \sim   \frac{i^l f_l}{4\pi} \cos(l\theta) \frac{l}{a r^2} $ thereby, exhibiting a $r^{-2}$ tail. Similarly, from (\ref{gq}) it is easy to show that for $l\neq 0$,
\begin{eqnarray}
\Delta \mathbf{j}_{l} (\mathbf{r}) &=& \hat{\mathbf{k}}_0 \left( i^l \frac{g_l}{4\pi} \cos(l\theta) \frac{r}{a^3} \sqrt{\pi} e^{-r^2 / 2a^2}\left[I_{\frac{l-1}{2}}\left(\frac{r^2}{2a^2}\right) - I_{\frac{l+1}{2}}\left(\frac{r^2}{2a^2}\right) \right] \right),
\end{eqnarray}
Therefore, the components of the current density too have the same behaviour, i.e., they go to zero as $r^l$ when $r<<a$ and have a $r^{-2}$ tail as $r\rightarrow \infty$.
\section{Calculation of $f_0, \, \text{and} \,  g_0$}\label{app2}
Converting the sum corresponding to (\ref{f2}) into an integral it can be shown that
\begin{eqnarray}\label{f3}
f^{+}(\theta_{\mathbf{q},\mathbf{k_0}}) &=& - \frac{1}{(2\pi)^2} \int_{0}^{2\pi}  d\theta_{\mathbf{k'}} \left[ \Big[V(\mathbf{0}) -\frac{1}{2}[1+ \cos(\theta_{\mathbf{k'}} -\theta_{\mathbf{k_0}})] V(\mathbf{k'-k_0})\Big]\frac{\hat{\mathbf{k'}}\cdot \hat{\mathbf{q}} \, k_{F}^{+}}{(\frac{1}{m}+\frac{\alpha}{k_{F}^{+}})(\mathbf{k'-k_0})\cdot \hat{\mathbf{q}}} \right], \nonumber \\
f^{-}(\theta_{\mathbf{q},\mathbf{k_0}}) &=& - \frac{1}{(2\pi)^2} \int_{0}^{2\pi}  d\theta_{\mathbf{k'}} \left[  \Big[V(\mathbf{0}) -\frac{1}{2}[1- \cos(\theta_{\mathbf{k'}} -\theta_{\mathbf{k_0}})] V(\mathbf{k'-k_0})\Big] \frac{\hat{\mathbf{k'}} \cdot \hat{\mathbf{q}}\,k_{F}^{-}}{\frac{1}{m}(\mathbf{k'-k_0})\cdot \mathbf{\hat{q}} - \alpha (\frac{\mathbf{k_0 \cdot \hat{q}}}{k_{F}^{+}}+\frac{\mathbf{k'\cdot \hat{q}}}{k_{F}^{-}})}  \right] . 
\end{eqnarray}
Next step is to consider the direction $\hat{\mathbf{k'}}$ to be fixed so that $\mathbf{k'}\cdot\hat{\mathbf{q}} = \cos \theta_{\mathbf{q}}$ and $\mathbf{k_0}\cdot\hat{\mathbf{q}} = \cos \theta_{\mathbf{k_0 ,q}}$ and determine $f_{l}^{s}$ using the following formula,
\begin{equation}\label{fs}
f_{l}^{s} = \frac{1}{\pi} \int_{0}^{2\pi} d\theta_{\mathbf{q}} \cos (l\theta_{\mathbf{q}}) f^{s}(\theta_{\mathbf{q},\mathbf{k_0}}), \, \text{with} \, s = \pm.
\end{equation}
\textit{Case 1, for $\alpha << v_0$ i.e., very small strength of RSOC}: Using (\ref{f3}) and the above equation, and evaluating the $\theta_{\mathbf{q}}$ integration first one arrive at the following expressions for $f_{0}^{s}$,
\begin{eqnarray}\label{f0}
f_{0}^{+} &=&-\frac{k_{F}^{+}}{4\pi^2 v_0}  \int_{0}^{2\pi}  d\theta_{\mathbf{k'}} \Bigg[V(\mathbf{0}) -\frac{1}{2}[1+ \cos(\theta_{\mathbf{k'}} -\theta_{\mathbf{k_0}})]  V( k_{F}^{+} \mathbf{\hat{k'}}- k_{F}^{+} \mathbf{\hat{k}_0})\Bigg] \nonumber \\
f_{0}^{-} &=& -\frac{k_{F}^{-}}{4\pi^2 v_0}  \int_{0}^{2\pi}  d\theta_{\mathbf{k'}} \Bigg[ V(\mathbf{0}) -\frac{1}{2}[1- \cos(\theta_{\mathbf{k'}} -\theta_{\mathbf{k_0}}) ]V( k_{F}^{-} \mathbf{\hat{k'}}- k_{F}^{+} \mathbf{\hat{k}_0}) \Bigg],
\end{eqnarray}
where $v_0$ is the degenerate Fermi velocity corresponding to both the sub-bands. This degeneracy is valid as long as the strength of the RSOC is small compared to the Fermi energy. In the above equations both $f_{0}^{+}< 0$, and $f_{0}^{-} < 0$ as is shown below for our chosen form of $V(q)$.  Let us now estimate the magnitude of the delocalized charge by evaluating the above integrations for our chosen form of electron-electron interactions. Using the explicit form of the potential $V(q)$ the quantity $f_{0}^{s} $ can be written as,
\begin{equation}\label{f0s}
f_{0}^{s}  = - \frac{k_{F}^{s}}{ v_0 \delta} +  \frac{k_{F}^{s}}{4\pi v_0} \int_{0}^{2\pi} d\theta_{\mathbf{k'}} \frac{[1 +s \cos(\theta_{\mathbf{k'}}-\theta_{\mathbf{k_0}})]}{\sqrt{(k_{F}^{s})^{2} +( k_{F}^{+})^{2} - 2k_{F}^{s}k_{F}^{+} \cos(\theta_{\mathbf{k'}}-\theta_{\mathbf{k_0}}) +\delta^{2}}}.
\end{equation}
Evaluating the above integral one can find the following expression for the magnitude of delocalization of charge resulting from the interactions coming from both the Rashba sub-bands, 
\begin{equation}\label{f0+}
f_{0}^{+} = - \frac{k_{F}^{+}}{v_0 \delta} +\frac{1}{4\pi v_0} \left[4 \sqrt{b+2} \left(K\left(\frac{2}{b+2}\right)-E\left(\frac{2}{b+2}\right)\right)\right]
\end{equation}
and 
\begin{equation}\label{f0-}
f_{0}^{-} = - \frac{k_{F}^{-}}{v_0 \delta} + \frac{k_{F}^{-}}{4\pi v_0\sqrt{(k_{F}^{-})^2 + (k_{F}^{+})^2 }} \left[4 \sqrt{b'} \left(E\left(-\frac{2}{b'}\right)-K\left(-\frac{2}{b'}\right)\right)\right]
\end{equation}
where $b =\frac{\delta^{2}}{2( k_{F}^{+})^{2}} $, and $b' =\frac{\delta^{2}}{(k_{F}^{-})^2 + (k_{F}^{+})^2 }$ and both $b$ and $b'$ are of $\mathcal{O}(r_s^{2})$, where $r_s$ is the dimensionless electron gas parameter. Here $K(z)$ and $E(z)$, with $z= \frac{2}{b+2}$ in the former and $z= \frac{2}{b'}$ in the later, are the complete elliptic integrals of first and second kind respectively \cite{AS}.  The quantity $v_0 f_0^{+}$ corresponding to (\ref{f0+}) is plotted as a function of the dimension less electron gas parameter $r_s$ in FIG. 6. In (\ref{f0+}), for very small $b$, the arguments of both the elliptic integrals approach to unity, i.e., $z \rightarrow  1^{-}$ and one can utilize  $z = 1- \epsilon $, with $\epsilon$ very small, to make series expansion of both $K(z)$ and $E(z)$ in the leading order in $\epsilon $ \cite{Wolf}. Then simplifications lead to,
\begin{equation}
f_{0}^{+} = - \frac{k_{F}^{+}}{v_0 \delta} - \frac{\sqrt{b+2}}{\pi v_0} \Bigg[ \frac{1}{2} \ln \Big(\frac{b}{b+2} \Big) + \frac{b}{b+2} \ln 2 +1 \Bigg],
\end{equation}
where $b \approx (r_s^{2})$ and $\frac{k_{F}^{s}}{\delta} \approx (r_s)^{-1}$. It can be easily seen from the above formula that the delocalized charge $f_{0}^{+}< 0$, and is inversely proportional to degenerate Fermi velocity $v_0 = \sqrt{\alpha^{2} + 2\mu / m}$.  For example, for InGaAs two-dimensional electron-gas (2DEG) with $r_s = 0.18 $ and $f_{0}^{+} \approx -\frac{5.44}{v_0}$.  On the other hand, one can replace $b' \approx r_s ^2$ for arbitrary values of $b'$ and $k_{f}^{-} / \delta$ by $r_s ^{-1}$ in (\ref{f0-}) to obtain,
\begin{equation}
f_{0}^{-} = - \frac{1}{v_0 r_s} + \frac{r_s}{ \pi v_0} \left[E\left(-\frac{2}{(r_s)^{2}}\right)-K\left(-\frac{2}{(r_s)^{2}}\right)\right]
\end{equation}
The above equation is plotted also in the FIG. 6, which shows $f_{0}^{-} <0$, for every values $r_s$. In the limit of very small  $b' \approx (r_s)^2$ using $E(-z) \approx \sqrt{z}$ and $K(-z) \approx \frac{ln(4\sqrt{z})}{\sqrt{z}}$, the equation (\ref{f0-}) turns out to be \cite{AS},
\begin{equation}
f_{0}^{-} = - \frac{k_{F}^{-}}{v_0 \delta} + \frac{k_{F}^{-}}{4\pi v_0\sqrt{(k_{F}^{-})^2 + (k_{F}^{+})^2 }} \left[4 \sqrt{b'} \left(\sqrt{\frac{2}{b'}}- \frac{\ln(4\sqrt{2/b'})}{\sqrt{2/b'}}\right)\right]
\end{equation}
which in the limit of $\alpha << v_0$ takes the form $f_{0}^{-} = -\frac{1}{r_s v_0} + \frac{1}{v_0}[0.45 - 0.4 r_s^2 + \ln (r_s)]$, and for InGaAs 2DEG we get $f_{0}^{-}\approx \frac{-6.83}{v_0}$. It is easy to recognize that the above expression of $f_{0}^{-}$ is always negative.  Furthermore, $f_{0}^{-} $ is inversely proportional to the $v_0$ and therefore,  depends on the strength $\alpha$ of the RSOC.
\begin{figure}[h]
\centering
\includegraphics[scale=0.45]{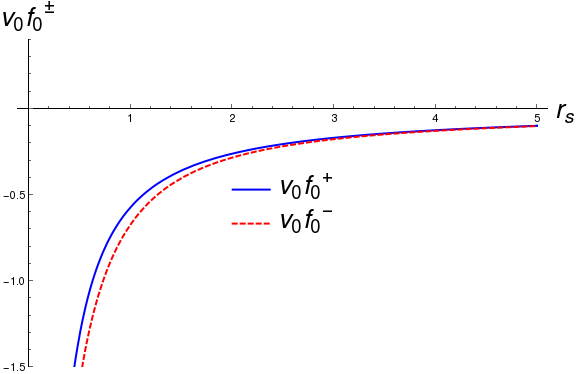}
\caption{$v_0 f_0^{\pm}$ as a function of the dimension less electron gas parameter $r_s$. In the limit of $r_s \rightarrow 0$, both the values of $v_0 f_0^+$ and $v_0 f_0^-$ diverge to $-\infty$, however in this limit of $r_s$ the electron density becomes too high and the system become a homogeneous electron gas (instead of being a Fermi liquid) and the Coulomb interaction become vary small compared to the kinetic energy of individual particles. Typical value of $r_s$ ranges \cite{Vignale} from 1 to 20 however (upto $r_s = 5$ is shown), even for InGaAs 2DEG with $r_s = 0.18 $ one finds $f_{0}^{+}\approx \frac{-5.4}{v_0}$ and $f_{0}^{-}\approx \frac{-6.83}{v_0}$ as mentioned above.}
\end{figure}
\paragraph*{}
Following the steps similar to that corresponding to the charge density, and using (\ref{jq+}) and (\ref{jso}) it can be easily shown that,
\begin{eqnarray}\label{gl}
\Delta \mathbf{j}(\mathbf{q}) &=& Q(\mathbf{q}) \left[\mathbf{g}^{+}(\theta_{\mathbf{q},\mathbf{k_0}}) + \mathbf{g}^{-} (\theta_{\mathbf{q},\mathbf{k_0}})\right] \nonumber \\ 
\nonumber \\
\text{where} \nonumber \\  \nonumber \\
\mathbf{g}^{s}(\theta_{\mathbf{q},\mathbf{k_0}}) &=& - \sum_{\mathbf{k'}} \Big( \frac{\mathbf{k'}}{m} + s \alpha \hat{\mathbf{k'}}\Big) \Big[V(\mathbf{0}) -\frac{1}{2}[1+ s \cos(\theta_{\mathbf{k'}} -\theta_{\mathbf{k_0}})] V(\mathbf{k'-k_0})\Big]\frac{\hat{\mathbf{k'}}\cdot \hat{\mathbf{q}} \, \delta(k'-k_{F}^{s})}{\frac{1}{m}(\mathbf{k'-k_0})\cdot \mathbf{\hat{q}} - \alpha (\mathbf{\hat{k_0}\cdot \hat{q}}-s \mathbf{\hat{k'}\cdot \hat{q}})} 
\end{eqnarray}
By fixing the $\hat{\mathbf{k}}_0$, i.e., the direction of propagation of the wave packet as the reference axis, the term $\cos(\theta_{\mathbf{k'}} -\theta_{\mathbf{k_0}})$ can be replaced by $\cos(\theta_{\mathbf{k'}})$ in the second one of the above equations. Then, converting the sum into integral over $d\mathbf{k'}$ and doing the $k'$ integration the above equation turns out to be,
\begin{eqnarray}
\mathbf{g}^{s}(\theta_{\mathbf{q},\mathbf{k_0}}) &=&
\int_{0}^{2\pi} d\theta_{\mathbf{k'}}\frac{ k_{F}^{s}}{4\pi} (\cos\theta_{\mathbf{k'}} \hat{x}+ \sin\theta_{\mathbf{k'}}\hat{y})  \frac{(1 +s \cos \theta_{\mathbf{k'}})}{\sqrt{(k_{F}^{s})^{2} +( k_{F}^{+})^{2} - 2k_{F}^{s}k_{F}^{+} \cos\theta_{\mathbf{k'}} +\delta^{2}}} \times \nonumber \\ & &
\frac{\cos\theta_{\mathbf{q}}}{(1-\cos\theta_{\mathbf{k_0}})\cos\theta_{\mathbf{q}}- \sin\theta_{\mathbf{k_0}} \sin\theta_{\mathbf{q}}}.
\end{eqnarray}
Then it is easy to determine the following,
\begin{equation}
g^{s}_0 \hat{\mathbf{k}}_0 = \int_{0}^{2\pi} d\theta_{\mathbf{k'}}\frac{ k_{F}^{s}}{4\pi}  \left(  \frac{\cos\theta_{\mathbf{k'}} \hat{x} (1 +s \cos \theta_{\mathbf{k'}})}{\sqrt{(k_{F}^{s})^{2} +( k_{F}^{+})^{2} - 2k_{F}^{s}k_{F}^{+} \cos\theta_{\mathbf{k'}} +\delta^{2}}} + \frac{ \sin\theta_{\mathbf{k'}}\hat{y}(1 +s \cos \theta_{\mathbf{k'}})}{\sqrt{(k_{F}^{s})^{2} +( k_{F}^{+})^{2} - 2k_{F}^{s}k_{F}^{+} \cos\theta_{\mathbf{k'}} +\delta^{2}}} \right)
\end{equation}
where the second integration vanishes as the integrand is an odd function, and we get the following expressions for $g_{0}^{+}$ and $g_{0}^{-}$ respectively,
\begin{eqnarray}
g_{0}^{+} &=& \frac{1}{2\sqrt{2}\pi} \int_{0}^{\pi} d\theta_{\mathbf{k'}} \left(  \frac{\cos\theta_{\mathbf{k'}}  (1 + \cos \theta_{\mathbf{k'}})}{\sqrt{1 +b -  \cos\theta_{\mathbf{k'}} }} \right) \nonumber \\
g_{0}^{-} &=& \frac{ k_{F}^{-}}{2\pi \sqrt{(k_{F}^{-})^2 + (k_{F}^{+})^2 }} \int_{0}^{\pi} d\theta_{\mathbf{k'}} \left(  \frac{\cos\theta_{\mathbf{k'}}  (1 - \cos \theta_{\mathbf{k'}})}{\sqrt{1 +b' -  \cos\theta_{\mathbf{k'}} }} \right) \nonumber \\
&\approx &  \frac{1}{2\sqrt{2}\pi} \int_{0}^{\pi} d\theta_{\mathbf{k'}} \left(  \frac{\cos\theta_{\mathbf{k'}} (1 - \cos \theta_{\mathbf{k'}})}{\sqrt{1 +b' -  \cos\theta_{\mathbf{k'}} }} \right)
\end{eqnarray}
where it is assumed that $\alpha << v_0$. If we assume that $b \approx b'$ then the value of $g_0 = g_{0}^{+}+ g_{0}^{-}$ can be found to be, 
\begin{equation}
g_0 = \frac{1}{\sqrt{2}\pi} \left[ \frac{ E(\frac{2}{b+2}) \left((b+1) \left[\frac{K\left(\frac{2}{b+2}\right)}{ E\left(\frac{2}{b+2}\right)}-1\right] -1\right)}{\sqrt{b+2}} \right].
\end{equation}
For any finite but small values of $b$ the argument of the complete elliptic integral $0<<z(=\frac{2}{b+2} )< 1$  and in this interval $K(z) >> E(z)$ \cite{AS}, thereby signifying $g_0 > 0$. On the other hand when one considers $b = 0$, corresponding to $r_s = 0$, one get a divergent $g_0$, however this limit corresponds to the high density electron gas and the Coulomb interaction does not operate and consequently the Fermi liquid picture is no longer required. It is worthwhile to point out that typical value of $r_s$ corresponding to 2D electron liquid (2DEL) ranges \cite{Vignale} from 1 to 20.
\paragraph*{}
\textit{Case 2, for strong RSOC:} In the case of strong SOC, when $m\alpha^2 /2 >> \mu >0$, the Fermi velocities corresponding to both the Rashba sub-bands no longer remain degenerate and the renormalized Fermi velocities follow, $\frac{v_\pm}{v_0} = 1 + \frac{1}{\pi v_0} \ln (\frac{k_F^{\pm}}{\delta})$, where $v_0 \approx \alpha$ and usually $\delta \sim k_{TF}$; $k_{TF}$ being the Thomas Fermi wave vector \cite{ARM}. However, the fact that quantity $\frac{k_{F}^{s}}{\delta}$ is $\mathcal{O} (r_s ^{-1})$ is enough for our purpose. Then from (\ref{f3}) and (\ref{fs}) it can be shown that,
\begin{eqnarray}\label{f02}
f_{0}^{+} &=&-\frac{k_{F}^{+}}{4\pi^2 v_+}  \int_{0}^{2\pi}  d\theta_{\mathbf{k'}} \Bigg[V(\mathbf{0}) -\frac{1}{2}[1+ \cos(\theta_{\mathbf{k'}} -\theta_{\mathbf{k_0}})]  V( k_{F}^{+} \mathbf{\hat{k'}}- k_{F}^{+} \mathbf{\hat{k}_0})\Bigg] \nonumber \\
f_{0}^{-} &=& -\frac{k_{F}^{-} (v_- - v_+ \cos\theta_{\mathbf{k_0}})}{4\pi^2 ((v_+)^2 +(v_-)^2 -2v_+ v_- \cos\theta_{\mathbf{k_0}})}  \int_{0}^{2\pi}  d\theta_{\mathbf{k'}} \Bigg[ V(\mathbf{0}) -\frac{1}{2}[1- \cos(\theta_{\mathbf{k'}} -\theta_{\mathbf{k_0}}) ]V( k_{F}^{-} \mathbf{\hat{k'}}- k_{F}^{+} \mathbf{\hat{k}_0}) \Bigg],
\end{eqnarray}
Rest of the calculations simply follow the steps which have been followed in obtaining the equations (\ref{f0s}), (\ref{f0+}) and (\ref{f0-}), and one can again show that both $f_0^{+}$ and $f_0^{-} <0$. From the above equations it is clear that $f_{0}^{-}|_{(\theta_{\mathbf{k_0}}= 0)} \propto -\frac{k_{F}^{-}}{4 \pi^2 (v_- - v_+)}$ when the c-QPWP is propagating along the `+ve' x-axis and $f_{0}^{-}|_{(\theta_{\mathbf{k_0}} = \pi)} \propto -\frac{k_{F}^{-}}{4 \pi^2 (v_- + v_+)}$ the c-QPWP is propagating along the `-ve' x-axis. Furthermore, $\frac{f_{0}^{-}|_{(\theta_{\mathbf{k_0}}= \pi)} }{ f_{0}^{-}|_{(\theta_{\mathbf{k_0}}= 0)} }\propto   \Big[\frac{ 2 \alpha \ln(2m\alpha^2 / \mu)}{ 2\pi \alpha + \ln(2m \alpha /\delta^2)} \Big]$. I have used the facts that in the strong RSOC \cite{ARM}, $k_{F}^{+} \approx \mu / \alpha << k_{F}^{-}\approx 2m\alpha$ and $v_0 \approx \alpha$. It is easy to recognize that for small $\delta(<< 2m \alpha)$ one gets $f_{0}^{-}|_{(\theta_{\mathbf{k_0}}= 0)} < f_{0}^{-}|_{(\theta_{\mathbf{k_0}}= \pi)}$ which indicates that in the case of strong RSOC, the magnitude of the delocalized charge depends quite strongly on the direction of  propagation of the c-QPWP. A similar analysis can be performed for the current density too and one finds,
\begin{eqnarray}\label{g02}
g_{0}^{+} &=& \frac{1}{2\sqrt{2}\pi} \int_{0}^{\pi} d\theta_{\mathbf{k'}} \left(  \frac{\cos\theta_{\mathbf{k'}} (1 + \cos \theta_{\mathbf{k'}})}{\sqrt{1 +b -  \cos\theta_{\mathbf{k'}} }} \right) \nonumber \\
g_{0}^{-} &=&  \frac{k_{F}^{-} (v_- - v_+ \cos\theta_{\mathbf{k_0}})}{2\pi \sqrt{(k_{F}^{-})^2 + (k_{F}^{+})^2} \left[(v_+)^2 +(v_-)^2 -2v_+ v_- \cos\theta_{\mathbf{k_0}}\right]} \int_{0}^{\pi} d\theta_{\mathbf{k'}} \left(  \frac{\cos\theta_{\mathbf{k'}} (1 - \cos \theta_{\mathbf{k'}})}{\sqrt{1 +b' -  \cos\theta_{\mathbf{k'}} }} \right) .
\end{eqnarray}
It can now be easily shown by following the steps corresponding to the \textit{Case 1} (corresponding to weak RSOC), that $g_0 >0$. However, it is easy to see from the above equations that the magnitude of both the delocalized charge and current strongly depend on the direction of propagation $\hat{\mathbf{k}}_0$, of the wave packet since the angle $\theta_{\mathbf{k_0}}$ determines the magnitudes in this case. The above equations further indicate that both the delocalized charge and current turn out to be non-trivial functions of the strength $\alpha$ of the RSOC.
\section{Estimation of volume over which charge is delocalized}\label{app3}
In this appendix, I estimate the 2D volume (area) over which the charge of a c-QPWP is delocalized. This is related to the finite time scale $\kappa ^{-1}$ (corresponding to the factor $e^{-\kappa t}$) of the `adiabatic switching on' of the interaction \cite{HK}. By the time the interaction is switched on, the chiral electrons with characteristic Fermi velocity $v_{\pm}$ reach a distance $R_{deloc} \sim v_{\pm} \kappa^{-1}$. Here the Fermi velocities $v_{\pm}$ correspond to the chiral quasi-particles with chirality $s= \pm 1$ respectively as mentioned earlier. The length scale $R_{deloc}$ represents the radius of the area over which the charge of the c-QPWP is delocalized. However, the time scale $\kappa^{-1}$ corresponding to the adiabatic switching on must be smaller than the lifetime $\tau$ of the quasi-particle, i.e., $\kappa ^{-1} \lesssim \tau$. This ensures $R_{deloc} \lesssim v_{\pm} \tau$, where $\tau ^{-1} = Im[\Sigma_s (\mathbf{k})]$. In the case of small RSOC corresponding to $v_{\pm} = v_0$, the $\tau ^{-1}$ turns out to be,
\begin{equation}
\tau ^{-1} \approx \frac{(\Delta k)^2}{2 \pi m} \left[ \frac{1}{2} + \frac{m \alpha ^2}{2 \mu} \ln \left(\frac{\alpha}{4}\sqrt{\frac{m}{2\mu}}\right) - \ln \left(\frac{\Delta k}{8\sqrt{2\mu m}}\right) \right],
\end{equation}
where $\Delta k = (k_0 - k_{F}^{+})$, and $\hbar = 1$ (here $k_F$ corresponding to Ref. [40] is identified here as $\sqrt{2\mu m}$ and $\delta$ of the same as $\delta k$)\cite{Sara}. Therefore,
\begin{equation}
R_{deloc} \lesssim 2 \pi m v_0 (\Delta k)^{-2} \left[ \frac{1}{2} + \frac{m \alpha ^2}{2 \mu} \ln \left(\frac{\alpha}{4}\sqrt{\frac{m}{2\mu}}\right) - \ln \left(\frac{\Delta k}{8\sqrt{2\mu m}}\right) \right]^{-1},
\end{equation}
which signifies that the quantity $R_{deloc}$ depends on the strength of the RSOC. The spread of the localized part of the c-QPWP is, on the other hand, given by $R_{loc} \sim (\Delta k)^{-1}$. Furthermore, it is easy to see that $R_{deloc}/R_{loc} >> 1$ in the limit of sufficiently small $\Delta k$ ($\rightarrow 0$) which is ensured by the choice of quasi-particles sufficiently close to Fermi surface. In this case one can make the length scale $R_{deloc}$ corresponding to the volume over which the charge of c-QPWP is delocalized arbitrarily larger than the length scale $R_{loc}$ corresponding to remaining localized charge (for the present case, represented by the Gaussian charge distribution corresponding to FIG. 3 (b)).
\end{appendix}
\end{widetext}


\begin{thebibliography}{100}
\bibitem{SO} In the crystalline enviornment when an electron with spin $\sigma$ moves it experiences an effective magnetic field $\mathbf{B}_{eff}$($ \sim \frac{\mathbf{p}}{m} \times \mathbf{E} $; with speed of light c=1) produced by the electric field ($\mathbf{E}$) corresponding to the gradiant of the crystal potential ($\mathbf{E} = -\nabla V_{crys}(\mathbf{r})$) \cite{Manc, Con}. The momentum dependent Zeeman energy of an electron moving under this effective magnetic field ($\mathbf{B}_{eff}$) is called the SO coupling. As a result electrons' spin ($\sigma$) and momentum ($\mathbf{p}$) degrees of freedom get coupled each other, and such systems possess electronic bands which are split by SO coupling.
\bibitem{RC} In this analysis, the coulomb interaction is considered to be of the Yukawa type form $V(r, \delta) \sim \frac{1}{r} e^{-\delta r}$ with characteristic length $\delta$ and electronic charge, $e = 1$, and it is operative in 3D, while the electrons are constrained to move in 2D. This gives, $V(\delta , |\mathbf{q}|)=\frac{2\pi}{\sqrt{|\mathbf{q}|^2 + \delta ^2}} $ in 2D (see (A 1.10) of  Ref. [41]). In Appendix \ref{app2}, I shall identify $\delta$ with Thomas-Fermi screening wave vector $k_{TF}$ for simplification purpose. In the case of a truely 2D coulomb interaction, being of the form $\ln(r)$, the Fermi liquid become unstable \cite{RC1}.
\end{thebibliography}
\end{document}